\documentclass[11pt]{article}
\usepackage{mathpazo}
\usepackage{eulervm}
\usepackage[T1]{fontenc}
\usepackage{inputenc}
\usepackage{geometry}
\usepackage{xcolor}
\usepackage{array}
\usepackage{verbatim}
\usepackage{float}
\usepackage{booktabs}
\usepackage{textcomp}
\usepackage{multirow}
\usepackage{amsmath}
\usepackage{amsthm}
\usepackage{amssymb}
\usepackage{graphicx}
\usepackage{rotfloat}
\usepackage{setspace}
\usepackage{lscape}
\usepackage{esint}
\usepackage[swedish,english]{babel}
\usepackage{caption, subcaption}
\usepackage[authoryear]{natbib}
\PassOptionsToPackage{normalem}{ulem}
\usepackage{ulem}
\usepackage{stackengine}
\usepackage[unicode=true,
 bookmarks=true,bookmarksnumbered=false,bookmarksopen=false,
 breaklinks=false,pdfborder={0 0 1},backref=page,colorlinks=false]
 {hyperref}

\definecolor{coolblack}{rgb}{0.0, 0.18, 0.39} 
\definecolor{cornellred}{rgb}{0.7, 0.11, 0.11}
\hypersetup{colorlinks=true,citecolor=coolblack, linkcolor=cornellred, urlcolor=coolblack} 

\makeatletter
\usepackage{color}
\usepackage{tikz,mathtools}
\usepackage{rotating}
\usepackage{threeparttable}
\usepackage{caption}

\definecolor{darkblue}{rgb}{0.0,0,.6}
\definecolor{maroon}{rgb}{0.68,0,0}
\definecolor{darkgreen}{rgb}{0,0.369,0.086}

\makeatother

\begin{document}
\title{\textbf{Duration Dependence and Job Search over the Spell: Evidence from Job Seeker Activity Reports}\thanks{We would like to thank Per-Anders Edin, Simon Ek, Peter Fredriksson, Axel Gottfries, Nikolaj Harmon, Jonas Maibom, Roland Rathelot and seminar participants at UCLS Annual Conference, IFAU, University of Edinburgh. Cederlöf gratefully acknowledges funding from the Swedish Research Council for Health, Working Life and Welfare (FORTE) Grant No. 2024-00417.}}
\author{Jonas Cederlöf\thanks{IFAU, University of Edinburgh and UCLS. Email: \href{mailto:jonas.cederlof@ifau.uu.se}{jonas.cederlof@ifau.uu.se}} \qquad Sara Roman\thanks{IFAU and UCLS. Email: \href{mailto:sara.roman@ifau.uu.se}{sara.roman@ifau.uu.se}}}
\date{November 2025}
\maketitle

\thispagestyle{empty}
\begin{abstract}

\noindent  We study how job search behavior evolves over the unemployment spell and the extent to which job seekers experience duration dependence in callbacks. Leveraging data on 2.4 million monthly \textit{activity reports} containing detailed information on job applications, interviews, and other search activities, we separate within-spell changes from dynamic selection with a time-and-spell fixed effects design. We find that raw search effort increases with unemployment duration, but this pattern reflects dynamic selection: within-spell search effort remains flat and declines sharply in the months preceding re-employment. Around unemployment insurance (UI) exhaustion, search effort drops by approximately 10\%, likely due to participation in labor market programs crowding out job search. Reported interviews indicate that callbacks decline by 6\% per month, but only 10--14\% of this decline reflects ``true'' duration dependence. Finally, we document substantial heterogeneity: search effort and duration dependence vary strongly by age, and job seekers in tight labor markets experience about 50\% more duration dependence. \\


\noindent \textbf{JEL-codes}: J64, J65, J68

\vfill{}
\end{abstract}
\pagebreak{}

\newpage{}
\pagenumbering{arabic}
\setcounter{page}{1}
\setstretch{1.3} 

\section{Introduction}
\label{sec:Introduction}

A well established empirical fact is that job-finding decreases with unemployment duration. One explanation for this pattern is dynamic selection, i.e that job seekers with lower \textit{ex ante} job-finding probabilities are more likely to remain unemployed. But to what extent other mechanisms such as reduced search effort\footnote{Throughout the paper, we use the terms search effort and search intensity interchangeably.} or unemployment duration itself having a causal effect on the job-finding rate -- commonly refereed to as structural or ``true'' duration dependence -- also contribute the decline, remains an open question. And, to what degree do these mechanisms vary across different types of job seekers and/or labor market conditions? Despite their importance for our understanding of the labor market and shaping of public policy, these questions remain poorly understood.

In this paper, we study how job search behavior evolves over the course of an unemployment spell and to what extent unemployed job seekers experience duration dependence. To explore this, we leverage data from 2.4 million \textit{activity reports}, submitted on a monthly basis by unemployed job seekers registered at the Swedish Public Employment Services (PES). These data cover the universe of UI-eligible job seekers and contain detailed information about all efforts undertaken towards securing a job, such as: applied-for jobs, attended interviews, participation in job fairs, training, education, sporadic employment and, in addition, the exact dates when each of these activities took place. Chiefly, the activities reported are not confined to any one platform, allowing for a holistic view of job seekers’ search behavior. The activity reports are also used by the PES to assess whether a job seeker meets the required job search criteria to maintain UI, thereby creating strong incentives for the unemployed to provide an accurate and comprehensive account of their efforts.

We begin by estimating a simple time-fixed effects model to see how search intensity evolves with time in unemployment for the stock of unemployed job seekers. At first glance, search effort appears to be increasing throughout the unemployment spell. However, this pattern arises solely due to dynamic selection, with job seekers applying for fewer jobs leaving unemployment at a higher rate. This \textit{prima facie} counter intuitive result reflects that job seekers with weaker labor market attachment need to apply for more jobs in order to secure employment. Once we account for dynamic selection using spell fixed effects, we find that within-spell search effort decreases throughout and within the unemployment spell. However, inspecting search intensity over realized durations, we see that search intensity actually remains largely constant within the spell and that the observed decrease in search effort is driven by a sharp drop in the number of applied-for jobs just before exiting unemployment; likely reflecting an interim period between having secured employment and starting the new job. Right-censoring the last two (calendar) months of unemployment and relating search effort to time until \textit{finding} a job, we find that search intensity is constant throughout the spell and estimates without spell fixed effects indicate that job seekers applying for more jobs leave unemployment at a higher rate. 

Next, we exploit variation in potential benefit duration (PBD) to examine how search effort evolves around UI exhaustion. We find that while remaining stable prior to exhaustion, it drops sharply -- by about 10\% -- at the point of exhaustion and then stabilizes at this lower level thereafter. About two thirds of the drop comes from job seekers no longer applying for any jobs (i.e., the extensive margin), whereas the remaining one third comes from lower search intensity on the intensive margin. At the same time, we observe an increase in other types of (potentially) employment enhancing activities, proportional to that of the drop in search effort. This pattern arises as unemployed job seekers become eligible for second-tier benefits --- generally at the same replacement rate as UI --- by participating in an active labor market program, which crowds out regular job search.\footnote{While the drop in search effort at UI exhaustion partly stands in contrast to previous literature -- where a decrease typically constitutes a reversion back to previous levels of search, having been preceded by spike, \citep[see e.g.][]{Marinescu2020,DellaVigna2021} -- we believe that the absence of a spike stems from UI exhaustion in the Swedish case not equating to benefit exhaustion. However, as acknowledged by \cite{DellaVigna2021}, the drop in search effort is also consistent with a model of job search with a fixed pool of vacancies where workers apply for the most available jobs before the point of exhaustion \citep[see][]{Faberman2019}.} 

We continue to leverage the richness of the data by exploiting its repeated structure, which includes job seekers’ reports of having attended multiple interviews throughout the spell along with the exact dates. First, we document that search effort is indeed predictive of future callbacks from employers, and has diminishing returns. On average, our estimates suggests that doubling the number of applied-for jobs over a three month period increases the probability of receiving a callback by about 30\%. Second, and in the spirit of correspondence studies examining differences callback rates across fictitious résumés with varying durations of non-employment \cite[e.g.][]{Kroft2013,Eriksson2014,Farber2016}, we isolate true duration dependence from that of the decline arising due to dynamic selection; again by using time and spell-fixed effects. Within the first year of unemployment, raw callback probabilities decline rapidly with time in unemployment at an average rate of about 5.6--6.3\% per month. However, accounting for dynamic selection attenuates the decline by almost an order of magnitude with callbacks decreasing by 0.76--0.87\%. These estimates remain robust to various alternative specifications and modeling choices. Our results thus point to a limited role of true duration dependence, suggesting that about 86\% of the observed decline in callback rates can be attributed to dynamic selection, leaving only 14\% due to true duration dependence. Although our estimates pertain to callbacks, this result closely aligns with recent evidence by \cite{Mueller2024} on duration dependence in job-finding rates. We show, in a simple stylized framework, that under certain conditions on the job-offer arrival rate --- consistent with recent empirical evidence \cite[see][]{Lalive2025} --- true duration dependence expressed as a share of the total observed decline, can approximately map to that in job finding, even in the presence of a potentially weak direct connection between callback rates and job-finding rates \citep{Jarosch2019}.

In the final part of the paper, we explore heterogeneity in job search behavior and duration dependence across different groups of job seekers and labor market conditions. While we observe meaningful differences in baseline levels of applied-for jobs, differences in how search intensity evolves over the unemployment spell are, for the most part, small and economically insignificant. For example, non-natives apply for about 27\% more jobs than natives at the outset of unemployment. Over the first year, natives reduce their search intensity by roughly 0.56\% per month, whereas non-natives apply for about 0.15\% more jobs per month. Where we do observe a pronounced difference in the evolution of search intensity is when we explore heterogeneity by age. Our analysis reveals substantial variation across the age distribution, exhibiting a clear increasing and concave relationship between age and search intensity: among the youngest job seekers search effort decreases by about 5\% per month, while for job seekers above age 38 the change in search effort turns positive and flattens out at around 1\% increase per month.

For callbacks, we find clear level differences across groups, but also, larger and likely more economically relevant effects on duration dependence and with a significant degree of heterogeneity. For example, non-natives receive about 29\% fewer callbacks and experience 17\% higher duration dependence compared to natives, lowering the probability of getting an interview by 0.95\% per month. Echoing the findings of \cite{Kroft2013} we also find that job seekers operating in tight labor markets also experience about 50\% greater decline in (within-spell) callback rates compared to job seekers in slack labor markets, lowering the probability of a callback by on average 1\% per month. We also find substantial heterogeneity in duration dependence across the age distribution, exhibiting an inverse U-shape. While the youngest job seekers experience a decline in callbacks by just about 2\% per month, the equivalent number is -0.6\% among middle-aged workers. The degree of duration dependence again starts to decline at age 59 and amounts to just about -2.2\% per month. 

We speak to several strands of literature. One focuses on job search behavior and its evolution throughout unemployment, either using data from online job boards \citep[e.g.][]{Baker2017,Marinescu2018,Faberman2019,Hensvik2021} or collected via (repeated) surveying of job seekers \cite{Krueger2011,DellaVigna2021}. A second closely related literature studies search effort in relation to UI exhaustion \cite[see e.g.][]{Marinescu2017,Marinescu2020,Lichter2021,DellaVigna2021}. A key advantage of our study compared to these litterateurs is the use of administrative data which alleviates concerns related to job seekers switching job search platforms or experiencing survey fatigue.\footnote{Of course, job seekers in our sample also may experience reporting fatigue, but, given non/misreporting leads to benefit sanctions, and on repeated offenses eventually having once benefits permanently revoked, we believe this to be less of a concern in our analysis.} A third strand of literature is the seminal work on duration dependence \citep{Ours1992,Berg1996} and in particular attempts at isolating the causal effect of unemployment duration on either measured by callbacks \citep{Kroft2013,Eriksson2014,Farber2016} or job-finding rates \citep{Heckman1984,Vandenberg2001,Alvarez2023,Mueller2024}. Our heterogeneity analysis also speaks the literature on differences in callback rates between workers with different ethnicity, gender and age \citep[see e.g.][]{Bertrand2004,Carlsson2007,Lahey2008,Neumark2019,Booth2012,Midtboen2016,Adermon2022}. Finally, we also relate to a recent literature that uses similar administrative records on individual job search behavior. \cite{Fluchtmann2024a,Fluchtmann2024b} study how average wages of applied-for jobs change with elapsed unemployment duration, and whether differences in job application behavior between men and women can account for part of the gender pay gap.

In concurrent work, \cite{Lalive2025} have digitized ``search diaries'' of 15,000 job seekers in Switzerland and, like us, study within-spell search effort and duration dependence in callbacks. One advantage of their data, compared to ours, is that it contains information on whether a specific job application resulted an interview, and eventually a job offer. This enables them to document how the likelihood of receiving a job offer, conditional on getting an interview, evolves over the course of an unemployment spell. A limitation, however, is that the timing of interviews and job offers are unknown, and that for about 40\% of job-applications these outcomes are not reported.\footnote{Whereas the probability of attending an interview among job seekers in our sample is just about 22\% at the beginning of the spell -- which is very similar to the callback rates found in the audit study by \cite{Eriksson2014} on Swedish job seekers -- the corresponding job seekers in the sample of Swiss job seekers in \cite{Lalive2025} is just about 5\%.} Because job seekers in their sample appear more likely to report complete information once an application is successful, interviews become rare events that typically occur near the end of the spell.\footnote{In our data, interviews are relatively frequent events with callback rates being largely constant within a spell (see Figure \ref{fig:interviewfixed} in Appendix \ref{app:appendixC}).} As a result, their setting prevents the use of spell fixed effects to estimate duration dependence in callback rates, requiring them instead to control for unobserved heterogeneity using a set of observed covariates. Our estimates on duration dependence in callbacks are thus arguably more robust to unobserved compositional differences arising over the spell.\footnote{\cite{Lalive2025} writes: ``the fact that we can only imperfectly account for unobserved heterogeneity leaves our empirical analysis inconclusive.'' (p. 3).}

Our study offers multiple contributions to the literature. First, we highlight how defining an unemployment spell in terms of time until \textit{finding} a job or time until \textit{starting} a job (i.e leaving unemployment) can have a substantial impact on conclusions about whether search effort declines or remains constant over the spell. We believe this observation can help clarify --- and potentially reconcile --- the conflicting evidence in the existing literature regarding the evolution of job search behavior throughout the duration of unemployment \cite[see e.g.][]{Krueger2011,Faberman2019,Marinescu2020,DellaVigna2021,Lalive2025}. Second, we document a drop in the \textit{level} of search effort at UI exhaustion; a somewhat rare finding that, to our knowledge, has only been reported by \cite{Krueger2011} for the US.\footnote{Other papers tend to find a decrease in search intensity which does not correspond to a level shift, but rather a reversion back to previous levels after having spiked at UI exhaustion \citep[see][]{Marinescu2020,DellaVigna2021}.} In our case, we argue that this drop arises because job seekers are required to participate in an active labor market program after UI exhaustion, which diverts time and effort away from job search. Third, we contribute to the literature on ``true''duration dependence by documenting a within-spell decline in callback rates and, using a simple stylized model, deriving the conditions under which these estimates are informative about the share of true duration dependence in job-finding rates. Alongside \cite{Lalive2025}, we are the first to document this pattern using administrative data on attended interviews, with the advantage that our data allow us to time interviews relative to unemployment duration and to address unobserved heterogeneity through spell fixed effects. Lastly, we provide some novel evidence of heterogeneity in both job search and duration dependence across job seekers and labor market conditions.

The paper unfolds as follows: In Section \ref{sec:institutional_setting_data} we describe the institutional details surrounding activity reporting as well as the data and the empirical strategy. In Section \ref{sec:evolution_of_search_behavior} we present results on how search effort evolves during the course of the unemployment spell and in relation to UI exhaustion. Section \ref{sec:duration_dependence} we estimate duration dependence in callbacks and separate between the part stemming from by dynamic selection and the causal effect of unemployment duration on individual callbacks and provide a simple framework for how these estimates could be informative about true duration dependence in job-finding. Section \ref{sec:heterogeneity_in_search_and_call_back_rates} explores heterogeneity in search effort and duration dependence across different groups of job seekers and labor market states. Finally, Section \ref{sec:conclusions} concludes.

\section{Institutional setting, data \& empirical strategy}
\label{sec:institutional_setting_data}

\subsection{Activity reporting, search requirements and monitoring} %
\label{sub:activity_reporting}

As of September 2013, unemployed job seekers registered at the Swedish Public Employment Service (PES) are required to submit monthly activity reports on their job-seeking activities.\footnote{This subsection draws heavily on information provided in official reports from The Swedish Unemployment Insurance Inspectorate \citep{IAF2015,IAF2016b,IAF2016,IAF2017,IAF2018a,IAF2018b,IAF2020,IAF2021,IAF2023}. } In these  reports, job seekers' must detail all efforts they have made in the preceding month to secure employment. This includes documenting: which jobs they applied for, interviews they attended, participation in training, education, job fairs or recruitment events, sporadic employment, or any other measures undertaken aiming towards increasing their employment prospects. The reports are very detailed and  require job seekers to specify the exact dates on which the activities took place.\footnote{For job applications and interviews the job seeker need also specify the type of position (occupation), the name of the employer, and in which city the job was located. Unfortunately, we do not observe this information in our data.} Figure \ref{fig:actreport} and \ref{fig:actreportonline} show the activity report templates for the paper and online version, respectively. 

The Swedish UI-system consists of two parts: The first part provides a low base amount which covers all workers with a sufficient work history. The second part is income-related which workers qualify for by having been a member in a UI-fund for 12 consecutive months prior to unemployment. Standard potential benefit duration (PBD) is 14 months (60 weeks), with an additional 7 months (30 weeks) granted to job seekers who at the time of benefit exhaustion have a child below the age of 18.\footnote{The income-related part of the system replaces 80 percent of a worker's previous wage subject to a benefit cap which is lowered after 20 weeks on UI. After 40 weeks of UI the replacement rate is lowered to 70 percent. Job seekers who exhaust their UI benefits can enroll in an active labor market program where they uphold \textit{activity support} at a 65\% replacement rate.} In order to uphold benefits during unemployment, job seekers need to fulfill a search requirement of actively be searching for suitable jobs. The activity reports serve as the basis for assessing whether the individual meets this requirement. Failure to submit a report or to fulfill the search requirement may result in sanctions in the form of suspended unemployment benefits and repeated offenses render benefits permanently revoked.\footnote{The sanctions follow a predetermined step-wise function: First offense renders a warning. A second offense results in the loss of one day's benefits, and a third or a fourth offense leads to five or ten days of suspended benefits, respectively. Job seekers committing a fifth offense, have their unemployment benefits permanently revoked and have to re-qualify through fulfilling a new employment condition.} 

During our period of study (2014--2019), the job search requirement was vaguely formulated as a job seeker having to be ``sufficiently active'' in their job search. The requirement was deliberately not anchored to a minimum number of job applications; instead, whether a job seeker was considered sufficiently active was determined on a case-by-case basis at the discretion of the individual's caseworker.\footnote{The Swedish Unemployment Insurance Inspectorate (IAF), supervisory authority overseeing the PES, have emphasized that these loosely defined requirements provided too much discretion to individual caseworkers, whom varied substantially in their leniency \cite{IAF2018a}.} As such, there was ambiguity and substantial uncertainty about what was levels of job search would suffice to avoid sanctions, not only among job seekers but also within the PES.\footnote{In a famous court case (Case 4864-19 ; 1526-20), the PES had upon request refused to provide a job seeker with a minimum level of applied-for jobs. Instead the PES emphasized that the number of jobs should be deemed ``sufficient''.} A vivid example of this is the PES reporting to their supervisory agency IAF, that in December of 2020 a job seeker could be considered sufficiently active if they had applied for one job in a month (IAF, 2021). However, about half a year earlier, legal precedence had been set in a court case, won by the PES, ruling that a job seeker was correctly sanctioned for having applied only to three jobs in one month (Case 1526-20).

Many caseworkers reported that reviewing activity reports was overly time-consuming and that following up on their specific content was often not a priority \citep{IAF2018b}. Although all activity reports were formally audited, several caseworkers acknowledged refraining from sanctioning job seekers they perceived as inactive, due to uncertainty about whether the job seekers had understood the requirements necessary to meet the minimum activity standards \citep{IAF2016}..\footnote{In an effort to clarify what was expected of job seekers, the PES introduced recommended targets on the number of applied-for jobs in February 2020 (outside of our sample period). These came in the form of intervals, with 6--20 jobs applied for per month being the standard recommendation. However, the minimum amount of applied-for jobs required to be considered active, was still determined on a case-by-case basis and not tied to the lower bound of the interval. As of November of 2022 the lower bound in the interval is a minimum requirement to be considered actively searching for jobs.} Because the search criteria were not anchored to a minimum number of job applications, and job seekers were unaware of how many applications would be needed to be considered “sufficiently active”, this uncertainty arguably provided incentives to report their job search activity accurately. As such, we believe that the number of job applications recorded in the activity reports largely reflects genuine search effort and represents, at least, a lower bound on actual job search activity.\footnote{This stands in contrast to \cite{Fluchtmann2024a} studying job search behavior among Danish job seekers. In their setting there is strong compliance to an explicit minimum search requirement of 1.5--2 jobs per week causing job seekers not to register jobs they have applied to in excess of the requirement, which is clearly indicated by an excess mass of job seekers reporting just hitting the target (see Figure 3 in \cite{Fluchtmann2024a} Online Appendix). As such, the authors conclude that in their setting ``[...] the number of applied-for jobs [is] uninformative about actual search behavior'' and they ``[...] therefore make no attempt to infer individual search effort from the observed number of applied-for jobs''\citep[][p. 1188]{Fluchtmann2024a}. Figure \ref{fig:hista} in Appendix \ref{app:appendixA} shows the distribution of the number of applied-for jobs in our sample, showing little sign of heaping at any specific number of applied-for jobs.} Nevertheless, we recognize that we cannot completely rule out the possibility that the patterns we see in data come from changes in reporting behavior as opposed to actual job search behavior.

\subsection{Data} 
\label{sub:data}
We make use of and combine multiple data sources. The main data source is the unemployment register, provided by the PES, containing all unemployed job seekers registered between 2000–2022. These data contain for each job seeker the start and end date of their unemployment spell(s) as well as detailed information about participation in labor market programs, subsidized employment, etc. Importantly, we also know from these data whether, and if so, when, a job seeker was deemed impaired and thereby exempted from reporting their activity.\footnote{This typically occurs when a worker has been deemed by their caseworker to have either a mental or physical impairment so severe that they cannot be considered to be ready to take up employment. In our data, 90\% of job seekers labeled impaired in any given month have not submitted an activity report.} We use these data to define an unemployment period (spell) as the time between registration and when the job seeker either de-registers from the PES or find part-time or subsidized employment lasting more than 30 days.\footnote{In Figure \ref{fig:earfoundjob} in Appendix \ref{app:appendixA} we confirm that ending an unemployment spells coincides with a sharp increase in the probability of employment and labor earnings.} These data are then coupled with data from the Swedish Unemployment Insurance Inspectorate (IAF) containing the universe of all UI-payments made to job seekers at the weekly level.

We add information from the monthly activity reports submitted by job seekers in 2014--2019. As described above, these reports contain self-reported information on, among other things, applied-for jobs and attended interviews in the previous month and the exact dates of these events. Finally, we link these data sources with multiple population registers from Statistics Sweden (SCB). Through the Longitudinal Integration Database for Health Insurance and Labor Market Studies (LISA), we obtain demographic information such as gender, age, country of birth, and education level of job seekers, as well as the age of their child(ren). Finally, we add matched employer-employee data (RAMS) containing information on annual earnings and monthly employment indicators covering the universe of the working population.

In our main analysis, we restrict the sample to job seekers who begin their unemployment spell as full-time unemployed and for whom at least one activity report is observed during the unemployment period. We further require that job seekers are eligible for unemployment benefits. Although activity reporting formally applies to all individuals registered with the PES, in practice it is only enforceable for those receiving unemployment benefits, as sanctions consist of withheld payments. We additionally limit the sample to new benefit periods in which job seekers are entitled to either 60 or 90 weeks of benefits, and where the first payment approximately coincides with the start of the unemployment spell. Finally, we exclude months during which job seekers are classified as impaired, since they are then exempt from activity reporting. After applying these restrictions, the final sample comprises 351,072 unemployment spells with 2,396,732 submitted activity reports, containing a total of 13,866,049 job applications and 868,240 attended interviews across 3,088,595 monthly observations

One difficulty lies in determining when exactly a job was found. Typically, employment begins some time after a job has been secured. Although individuals who have secured employment are, in principle, expected to continue searching for jobs until their new employment starts, it is not obvious that this interim period is of any great interest as the incentives to apply for additional are arguably small. Foreshadowing the results of Section \ref{sec:evolution_of_search_behavior}, we see a distinct drop in the number of applied-for jobs during the last two months of unemployment. Whereas the drop in the very last month is largely an artifact of measuring activity in calendar months, we interpret the sharp decrease preceding it as reflecting job seekers having secured employment and are awaiting its start.\footnote{A mechanical calendar-month effect is also visible for the first month of unemployment. In other words, job seekers entering (exiting) their spell late (early) on in a month, will by construction have fewer jobs applied for in that month.} Similar to \cite{Fluchtmann2024a}, we therefore censor the last two calendar months of each unemployment spell for the main part of our analysis.\footnote{\cite{Fluchtmann2024a,Fluchtmann2024b} imposes a similar restriction, considering the last four weeks as the interim period between finding a job and starting it, thereby dropping them from the analysis. As the day-of-the-month of exiting unemployment is uniformly distributed in our sample, our restriction implies dropping the last 6 weeks of a spell on average.} As a result, we exclusively study unemployed individuals whose unemployment duration exceeds at least  two (calendar) months. In Section \ref{sub:search_intensity_until_finding_a_job}, we show that this restriction is pivotal in shaping the estimated evolution of search effort over the unemployment spell. We also provide additional empirical justification for censoring the last two months, which yields a more accurate representation of actual search behavior within unemployment spells.

\subsection{Descriptive statistics}

\begin{table}[t]\centering \small
\begin{threeparttable}[t] 
\caption{Descriptive statistics} \label{tab:desc}
\def\sym#1{\ifmmode^{#1}\else\(^{#1}\)\fi}  \begin{tabular}{l*{10}{c}} 
\toprule
& All UI-eligible & \multicolumn{3}{c}{Distribution of elapsed duration} \\
& job seekers & $<p25$ & $p25-p75$ & $>p75$ \\
\midrule
\textbf{Demographics} \\ Age&       40.6&       38.1&       40.4&       43.5\\
Female              &        0.45&        0.51&        0.44&        0.41\\
Non-native       &        0.35&        0.25&        0.33&        0.50\\
Married             &        0.36&        0.32&        0.35&        0.42\\
Child (aged 0–16) &        0.34&        0.35&        0.33&        0.33\\
Monthly earnings ($t-1$) & 21,536&    21,884&    22,125&    20,013\\
Level of Education \\ \quad Primary&        0.14&        0.09&        0.13&        0.22\\
\quad Upper secondary&        0.51&        0.53&        0.52&        0.47\\
\quad Post secondary&        0.35&        0.38&        0.35&        0.31\\
\addlinespace
\textbf{Spell characteristics} \\
Unemployment duration (days)&      304&       61&      215&      729\\
\# activity reports &        6.83&        1.69&        4.99&       15.68\\
\# applied-for jobs &       39.50&        8.06&       28.12&       93.92\\
\# interviews       &        2.47&        1.23&        2.57&        3.54\\
\midrule 
\# individuals&   329,622&    83,544&   163,906&    82,172\\
\# spells           &   351,072&    88,481&   174,910&    87,681\\
\# observations                   &  3,088,595&   238,359&  1,189,276&  1,660,960\\
\bottomrule 
\end{tabular} 
\begin{tablenotes}[flushleft] 
\scriptsize \item \noindent Notes: The table shows sample averages at the spell level.
\end{tablenotes} 
\end{threeparttable} 
\end{table}

\label{sub:descriptive_statistics}
Column (1) of Table \ref{tab:desc} shows job seeker characteristics for all unemployment spells in our sample (not yet having right-censored the last two months of the spell). On average, job seekers are 40.7 years old starting the unemployment, just over 44\% are women and 32\% are foreign-born. The vast majority of job seekers have completed upper secondary education, while 14\% have completed primary education. Overall, our sample is relatively representative of the full population of full-time unemployed registered with the PES, except for being more established in the labor market which follows from restricting attention to job seekers entitled to unemployment benefits.

Columns (2)–(4) break down our sample based job seekers position in the unemployment-duration distribution. Here, it becomes apparent that dynamic selection occurs throughout the spell, i.e. that job seekers with longer periods of unemployment are adversely selected in terms of their characteristics. For example, we see that the proportion of foreign-born individuals increases among the unemployed as the unemployment period progresses. Among those in the first quartile of unemployment duration (Column 2), 25\% are foreign-born, whereas in the fourth quartile (Column 4), where unemployment lasts over 2 years, the proportion rises to 49\%. The same pattern is observed among individuals whose highest level of education is primary school, with the share increasing from 14\% in the first quartile to 22\% in the fourth quartile.

This type of dynamics selection implies that depending on when in the unemployment period we study job search behavior, we will be observing different types of individuals. In the next section, we describe in more detail how this affects the interpretation of our results and how we go about accounting for this dynamic selection. 

\subsection{Empirical strategy} 
\label{sub:empirical_strategy}

We aim to describe how job seekers' search behavior and interview (callback) probabilities evolve throughout the unemployment spell. To this end, we start by estimating the following model:
\begin{equation}
\label{eq:timeFE}
	y_{it} = \tau_{t} + \varepsilon_{it}
\end{equation}
where $y_{it}$ represents job search behavior (i.e the number of applied-for jobs or attended interviews) for individual $i$ in month $t$ relative to the start of unemployment. $\tau_t$ is a set of time-fixed effects showing the average number of applied-for jobs or interviews at a given duration. Lastly, $\varepsilon_{it}$ is the error term capturing individual deviations from the mean.

As noted above, the estimates of $\tau_t$ will inevitably be subject to dynamic selection. That is, if individuals who apply for more jobs exit unemployment earlier than those who apply for fewer jobs, the latter group will make up a larger portion of the sample on which the average is calculated. Thus, $\tau_t$ reflects the average number of job applications (or interviews) among those who remain unemployed. However, these estimates contain little information about how a given individual (on average) changes their job search behavior during the unemployment spell. In other words, the estimates may to a large degree be driven by compositional changes in the sample.

To capture how job seekers change their job search behavior within the unemployment spell, we control for compositional changes in the sample by estimating the model:
\begin{equation}
\label{eq:spellFE}
	y_{it} = \alpha_i + \tau_{t} + \varepsilon_{it}
\end{equation}

\noindent where $\alpha_i$ are spell-fixed effects. These fixed effects control for level differences in the number of applied-for jobs (interview probabilities) across job seekers. In this way, we control for dynamic selection arising due to some job seekers, on average, applying for more jobs and therefore may exit unemployment earlier.

The model in equation \eqref{eq:spellFE} accounts for dynamic selection in levels across spells. However, it does not address potential differences in search dynamics—specifically, that some individuals may increase or decrease their job search effort more than others and therefore exit unemployment earlier or later. To interpret $\hat{\tau}_t$ as capturing changes in job search behavior for a given individual, one must assume that all individuals adjust their effort in the same way over time \citep{Fluchtmann2024a}. In the following section, we follow \citet{Fluchtmann2024a} and show that this assumption is likely satisfied, indicating an absence of dynamic selection in changes and supporting an interpretation of the $\hat{\tau}_t$ estimates from equation \eqref{eq:spellFE} as reflecting within-spell changes in job search behavior over the course of the unemployment spell.

\section{Job search behavior over the unemployment spell}
\label{sec:evolution_of_search_behavior}
In this section we present results on how overall search intensity evolves with unemployment duration and exploit differences in PBD to study how search effort changes around UI exhaustion.

\subsection{Search intensity during unemployment} 
\label{sub:search_intensity}

\begin{figure}[t]
  \caption{Number of applied-for jobs by unemployment duration} 
  \label{fig:searchallmonths}
 \begin{subfigure}[a]{\textwidth}
 \centering 
 \includegraphics[width=.65\textwidth]{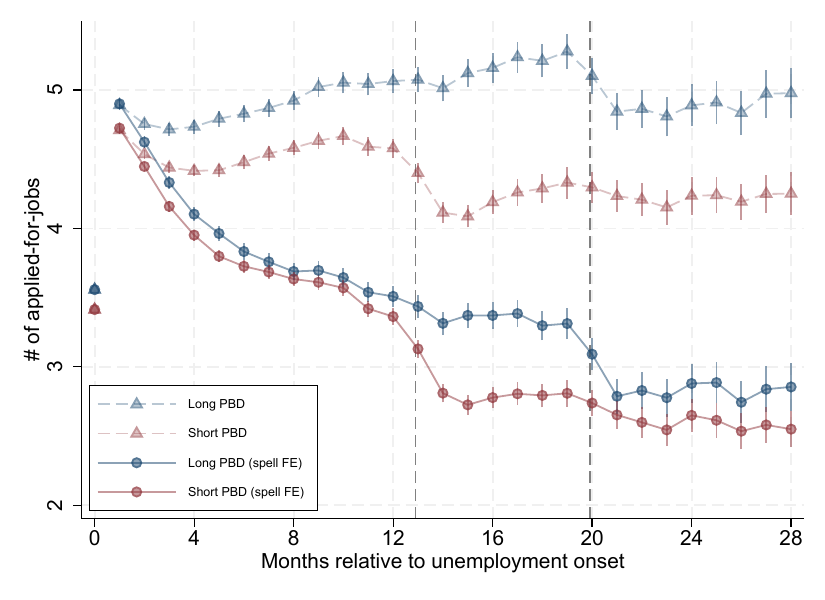}
\end{subfigure}
\footnotesize Notes: The figure shows the evolution of search intensity relative to unemployment onset for job seekers entitled to 14 (red) or 21 (blue) months of UI indicated by the dashed vertical lines. The circles connected by dashed lines show estimates from equation \eqref{eq:timeFE} and reflect average number of applied-for jobs among the sample of job seekers remaining in unemployment in a given month. The triangles connected by solid lines depict estimates from \eqref{eq:spellFE}, accounting for dynamic selection and reflect within-spell changes in search intensity. Surrounding each estimate are 95\% confidence intervals with standard errors are clustered at the spell level.
\end{figure}

Figure \ref{fig:searchallmonths} shows average number of applied-for jobs by time in unemployment separately for job seekers eligible to 14 (red) and 21 months (blue) of PBD. The triangles in lighter colors which are connected with dashed lines show estimates from equation \eqref{eq:timeFE}, representing averages among the stock of job seekers still unemployed at a given point in time. The darker colored circles, connected with solid lines, depict estimates from equation \eqref{eq:spellFE}, showing the within-spell evolution of search effort.

After an initial phase in period, search effort peaks at around 4.7--4.9 applied-for jobs in the month following onset of unemployment. After a slight decrease, lasting up until month 4 of unemployment, number of applied-for jobs grows steadily with unemployment duration up until the point of UI exhaustion where it drops by about 10\%. We will return to this observation and elaborate upon the drop in search effort at UI exhaustion in Section \ref{sub:search_intensity_until_finding_a_job}. Importantly, these estimates reflect search intensity among the sample of job seekers remaining in unemployment at each duration and are thus subject compositional changes over time (i.e dynamic selection). Once we control for dynamic selection by including spell-fixed effects within-spell search effort is in fact decreasing during the first year of unemployment and exhibits a similar drop at UI exhaustion, thereafter remaining largely constant. Taken together, this implies that unemployed job seekers are positively selected over time in terms of their search effort. That is, individuals with greater search intensity are those who tend to be unemployed for longer periods. This seemingly counter intuitive fact reflects that job seekers with \textit{ex ante} lower labor market prospects have to apply for more jobs in order to escape unemployment.

\begin{figure}[t]
  \caption{Number of applied-for jobs by fixed durations}
    \label{fig:fixeddur}
     \begin{subfigure}[a]{.49\textwidth}
     \centering 
    \caption{Short PBD}
     \label{fig:fixeddura}
      \includegraphics[width=\textwidth]{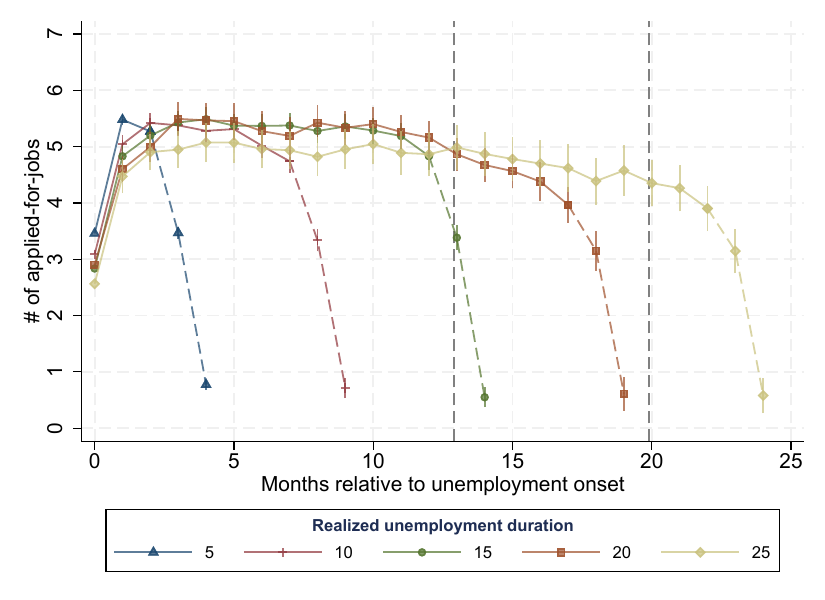}
    \end{subfigure}
    \begin{subfigure}[a]{.49\textwidth}
     \centering
    \caption{Long PBD}
    \label{fig:fixeddurb}
      \includegraphics[width=\textwidth]{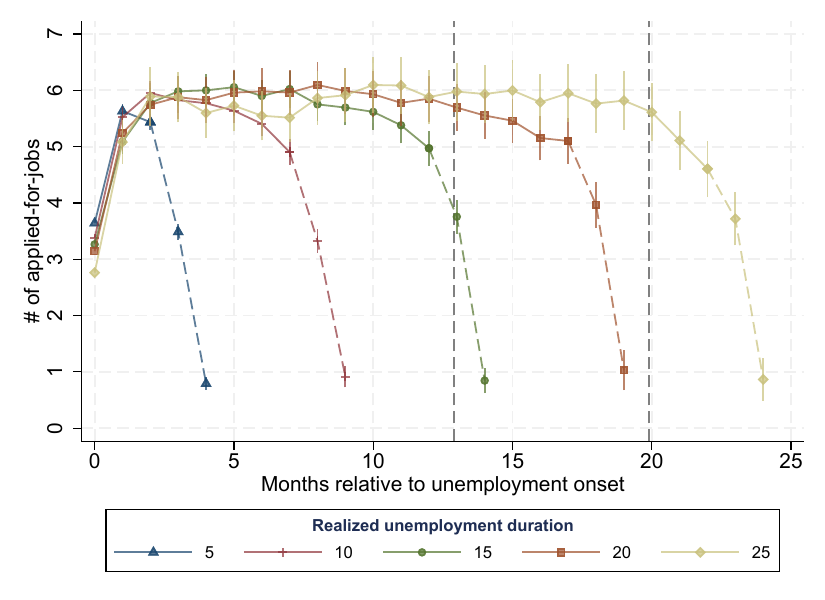}
    \end{subfigure}
    \footnotesize Notes: The figure shows the evolution of search intensity relative to unemployment onset for job seekers with 5, 10, 15, 20 or 25 months of realized unemployment duration estimated using the model in equation \eqref{eq:timeFE}. Panel a) plots job search intensity for job seekers entitled to 14 months of UI and panel b) for job seeker entitled to 21 months. The dashed portion of each solid line reflect the last two calendar months in the unemployment spell which we view as being the interim period between securing employment and starting a new job. Surrounding each estimate are 95\% confidence intervals with standard errors are clustered at the spell level.
\end{figure}

To investigate how well the spell-fixed effects model fits the data and whether dynamic selection arising from differences in search dynamics over time influence our estimates, Figure \ref{fig:fixeddur} plots the average number of applied-for jobs by realized duration. We again separate between job seekers with short and longer PBD in Figure \ref{fig:fixeddura} and \ref{fig:fixeddurb}, respectively. We make two observations from these graphs: First, the dynamic pattern of search looks very similar across job seekers with different realized unemployment durations. This suggests that there is little evidence for any dynamic selection in changes occurring such that e.g. job seekers with longer durations have a more rapid decline in their search effort and therefore leave unemployment later. As such, we believe that the spell-fixed model can deliver an accurate description of how search effort evolves throughout the unemployment spell. Second, consistent with other papers measuring search effort in terms of time spent on job search \citep[c.f.][]{Krueger2011,DellaVigna2021}, search effort appears relatively stable within the unemployment spell. However, there is a noticeable drop in the number of applied-for jobs during the last two months of the spell. As described in section \ref{sub:data}, part of this is mechanical and due to spells ending early in a calendar month. Nevertheless, the drop in search effort is also apparent in the second to last month of the spell. We interpret this drop as stemming from job seekers having landed a job and await its start. This interpretation is corroborated by the fact that we see the equivalent pattern, and even more pronounced, when we plot the probability of attending an interview by realized durations (see Figure \ref{fig:interviewfixed}). We probe this interpretation further by plotting employment and earnings by months relative to job seekers last recorded interview in the unemployment spell (see Figure \ref{fig:earlastinter}). This shows both earnings and employment rates are largely constant in the months preceding the last interview and increasing drastically just two calendar months after it. This points to two calendar months being a reasonable approximation of the interim period between having found a job and starting it. 

In conclusion, we view Figure \ref{fig:fixeddur} as portraying a largely constant search intensity within the unemployment spell. Once we get rid of the mechanical effect and drop only the last (calendar) month of unemployment we see a slight decrease in search effort over time (shown in Figure \ref{fig:searchback0a} in Appendix \ref{app:appendixB}). Part of this decrease, however, still stems from job seekers having secured employment and are awaiting its start. This interim period is arguably of less interest, in particular since the incentives to apply for new jobs decreases substantially. Hence going forward, we right-censor the last two months of each unemployment spell and consider search effort in relation to time until \textit{finding} a job.

\subsection{Search intensity until finding a job and UI exhaustion} 
\label{sub:search_intensity_until_finding_a_job}

\begin{figure}[t]
  \caption{Number of applied-for jobs by time until finding a job} 
  \label{fig:search}
 \begin{subfigure}[a]{\textwidth}
 \centering 
 \includegraphics[width=.65\textwidth]{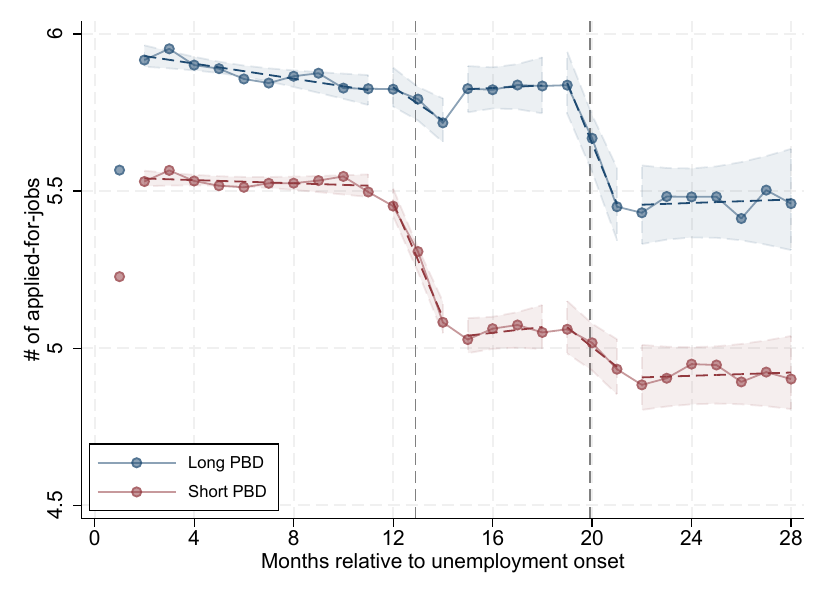}
\end{subfigure}
\footnotesize Notes: The figure shows within spell variation in the number of applied-for jobs by unemployment duration. The blue and red dots show estimates from equation \ref{eq:spellFE} run separately for job seekers eligible to 21 and 14 months of PBD on UI, respectively, which are indicated by the vertical dashed line. The colored dashed lines show linear predictions along with 95\% confidence bands, having regressed number of applied-for jobs on elapsed duration, interacted with an indicator for different segments $k$ of duration $t$ being within the interval $k=\{[2,11],[12,14],[15,18],[19,21],[22,28]\}$. All spells are right-censored at the last two calendar months of each unemployment spell. Standard errors are clustered at the spell level.

\end{figure}

Figure \ref{fig:search} plots estimates from equation \eqref{eq:spellFE} on the number of applied-for jobs by elapsed duration, right-censoring the last two calendar months in the unemployment spell (6 weeks on average). We again separate between job seekers with short (red) and long (blue) PBD. As a summary measure on the evolution of search, the dashed colored lines show linear predictions, along with 95\% confidence bands, from regressing the number of applied-for jobs on elapsed duration, interacted in different segments.\footnote{Specifically, we estimate the following regression:
\[
y_{it} = \alpha_i + \sum_{k} \left[ \gamma_k (D_k \times t) + \delta_k D_k \right] + \varepsilon_{it}
\]
where $t$ denotes months in unemployment, and $D_k$ is an indicator  equal to one if $t$ falls within interval $k \in \{[2,11], [12,14], [15,18], [19,21], [22,28]\}$ and 0 otherwise.}  The regression estimates are presented in Table \ref{tab:mainsearch} along with the implied average monthly percentage change which we calculate with compound decay.
\footnote{We calculate the average monthly percentage change within each interval \(k\) using a compound decay rate formula: $
\Delta_k\% = 1 - \left( \frac{\hat{y}_{\overline{k}}}{\hat{y}_{\underline{k}}} \right)^{\frac{1}{\overline{k} - \underline{k}}},$
where \( \hat{y}_{\underline{k}} \) and \( \hat{y}_{\overline{k}} \) denote the predicted outcomes at the beginning and end of interval 
\( k \in[\underline{k},\overline{k}] \).}

\begin{table}[t]\centering 
\begin{threeparttable}[b]
\def\sym#1{\ifmmode^{#1}\else\(^{#1}\)\fi} \small
\caption{Search effort by elapsed duration} \label{tab:mainsearch}
\begin{tabular}{l*{4}{c}}
\toprule
&\multicolumn{2}{c}{Short PBD}&\multicolumn{2}{c}{Long PBD} \\
\cmidrule(lr){2-3} \cmidrule(lr){4-5}  
&\multicolumn{4}{c}{applied-for jobs per month}\\
& $ \Delta $\# & $ \Delta $\%  & $ \Delta $\# & $ \Delta $\%  \\
&(1) & (2) & (3) & (4) \\
\midrule 
\multicolumn{2}{l}{Duration (months)}&&\\
 
\quad 2--11 &      -0.003         &      -0.05         &      -0.012\sym{***}&      -0.21         \\
            &     (0.003)         &                     &     (0.004)         &                     \\
 
\quad 12--14&      -0.183\sym{***}&      -3.40         &      -0.053\sym{**} &      -0.92         \\
            &     (0.015)         &                     &     (0.021)         &                     \\
 
\quad 15--18&       0.011         &       0.22         &       0.006         &       0.11         \\
            &     (0.013)         &                     &     (0.017)         &                     \\
 
\quad 19--21&      -0.059\sym{**} &      -1.18        &      -0.194\sym{***}&      -3.38         \\
            &     (0.023)         &                     &     (0.031)         &                     \\
 
\quad 22--28&       0.003         &       0.05         &       0.004         &       0.07         \\
            &     (0.011)         &                     &     (0.016)         &                     \\
\cmidrule(lr){2-3} \cmidrule(lr){4-5} 
\# applied-for jobs ($ t$=2) &\multicolumn{2}{c}{5.54}&\multicolumn{2}{c}{5.93}\\
\# observations&\multicolumn{2}{c}{1,565,787}&\multicolumn{2}{c}{785,113}\\
\bottomrule
\end{tabular}
\begin{tablenotes}[flushleft]
\scriptsize \item \noindent  Notes: The table shows in columns (1) and (3) the average within-spell change in the number of applied-for jobs per month by different segments of elapsed duration. The estimates come from the regression: $ y_{it} = \alpha_i + \sum_{k} \left[ \gamma_k (D_k \times t) + \delta_k D_k \right] + \varepsilon_{it}$ where $t$ denotes months in unemployment, and $D_k$ is an indicator  equal to one if $t$ falls within interval $k \in \{[2,11], [12,14], [15,18], [19,21], [22,28]\}$ and 0 otherwise. Columns (2) and (4) display the implied average percentage change per month calculated by plugging in the predictions from the ends of each interval $k\in[\underline{k},\overline{k}]$ into the compound decay formula: $
\Delta_k\% = 1 - \left(\hat{y}_{\overline{k}}/\hat{y}_{\underline{k}} \right)^{\frac{1}{\overline{k} - \underline{k}}}$. All spells are right-censored at the last two calendar months of each unemployment spell. Standard errors are in parenthesis and clustered at the spell level. Asterisks indicate that the estimates are significantly different from zero at the  \sym{*} \(p<0.1\), \sym{**} \(p<0.05\), \sym{***}  \(p<0.01\) level.
\end{tablenotes}
\end{threeparttable}
\end{table}

Considering time until finding a job, search intensity turns out to be largely constant throughout the unemployment spell.\footnote{Figure \ref{fig:searchback0b} in the Appendix \ref{app:appendixB} shows that the difference in levels between the estimates with and without spell-fixed effects. Comparing these estimates when right-censoring the last two months of the spell now suggest negative selection on search intensity over time. I.e that job seekers who are applying to more jobs secure employment faster.} For job seekers with short PBD the change in search intensity is not significantly different from zero within the first year of unemployment whereas for job seekers with long PBD, search intensity decreases by merely 0.2\% per month. 

There is a significant drop in search intensity occurring around each PBD-groups respective UI exhaustion point, amounting to 3.4\% per month.\footnote{There is also a significant, although smaller sized drop in search intensity around the ``wrong'' UI exhaustion point for each PBD-group. These likely originate from children having turned 18 at exhaustion due to the job seeker not having utilized 5 days of UI per week and that we are only able to identify the year of birth of the youngest child.} Thus, around the period of UI exhaustion, search intensity drops by just about 10\% in total. This result partly contrasts earlier research of search effort around UI exhaustion. While \cite{Marinescu2020} and \cite{DellaVigna2021} both find a drop in search effort following UI exhaustion, the drop constitutes a reversion back to previous levels, which have been preceded by a sharp rise (spike) just at UI exhaustion.\footnote{Consistent with much of the previous literature \cite[see, e.g.][]{Katz1990,Schmieder2012,Marinescu2020,DellaVigna2021}, we do see a spike in the hazard just around UI exhaustion for each PBD group (see Figure \ref{fig:hazard} in Appendix).} Our results suggest more of a downward shift in search effort following UI exhaustion, with the change in number of applied-for jobs being largely constant and insignificant both before and after. To understand this discrepancy in results we first note that, in Sweden, unemployed job seekers who exhaust their UI benefits are offered to participate in an active labor market program (the \textit{Job and Development Guarantee}) which consists of job search assistance, counseling, job training and practice, and vocational training. Participation in the program qualifies job seekers for alternative benefits (\textit{activity support}) at slightly lower replacement rate compared to UI.\footnote{The nominal replacement rate for activity support is 5\% lower compared to UI. However, in practice the net replacement rate change in our sample is -1.45\% as a large fraction have previous earnings above the benefit cap.} Not enrolling in the program means having to rely on household-level, means-tested social assistance, providing significantly lower benefit levels. Therefore, the vast majority of job seekers remaining unemployed after UI exhaustion do enroll \citep[see][]{Cederlof2025a}. Importantly, in order to uphold activity support, job seekers have to continue reporting their search efforts by submitting monthly activity reports. We do see that about two thirds of the drop in search intensity at exhaustion comes from job seekers failing to submit their activity report. However, given that the incentives for reporting have not changed substantially, we are inclined to interpret this as job seekers actually lowering their search intensity, as opposed to only lowering their reporting propensity. 

In Appendix \ref{app:appendixB} we provide evidence corroborating this interpretation by studying whether job seekers report other types of activities and by studying search intensity among job seekers having submitted an activity report Table \ref{tab:otherint} and Figure \ref{fig:otherint}) show a 2.5--3.5\% monthly increase in the probability of participating in other types of activities around UI exhaustion. Unfortunately, we lack detailed information on the exact nature of the activities, but they likely include education, job practice and vocational training assigned by caseworkers within the JDG. Among job seekers submitting an activity report, we see a drop in search effort around UI exhaustion by 0.9\% and 1.2\% decrease per month for job seekers with long and short PBD, respectively. These estimates reflect a lower bound of the full drop in search effort as we believe a non-negligible share of job seekers failing to submit a report also reflect not having applied for any jobs. In conclusion, we find that UI exhaustion is coupled with a decrease in search intensity ranging between 3--10\% and that this drop likely occurs due to participation in active labor market programs crowding out regular job search.

\subsection{Returns to search effort}
\label{sub:marginal_return_to_search_effort}
We now turn to estimating the returns to job-search effort. To this end, we draw on job seekers' self-reported interview attendance and estimate the following fixed-effects regression
\begin{equation}
	\label{eq:searchreturn}
	\text{Pr}(Interview)_i = \alpha_i + \tau_t + \sum_{k=-4}^{4} \phi_{t+k} \text{log}(Search \, effort + 1)_{i(t+k)} + \nu_{it}. 
\end{equation}
Here, search effort is measured as the number of job applications submitted.\footnote{While modeling search effort using $\log(x+1)$ is not ideal \cite[see][]{Chen2023}, estimating the model in levels with quadratic terms and evaluating the marginal effect at the mean yields similar results.} The coefficients of interest, $\phi_{t+k}$, capture the semi-elasticity with respect to search effort --- that is, the percentage point change in the probability of being called to an interview in month $t$ resulting from a 1\% increase in job applications in month $t+k$, holding search effort in surrounding months constant. By including both time and spell fixed effects, the regression controls for unobserved, time-invariant heterogeneity across job seekers as well as general duration dependence in callback probabilities. For ease of interpretation, we estimate the returns to search by pooling PBD-groups and focus on the first 12 months of unemployment.

\begin{figure}[t]
  \caption{Returns to search effort}
    \label{fig:searchreturn}
     \begin{subfigure}[a]{.49\textwidth}
     \centering 
    \caption{Dynamics}
     \label{fig:searchreturna}
      \includegraphics[width=\textwidth]{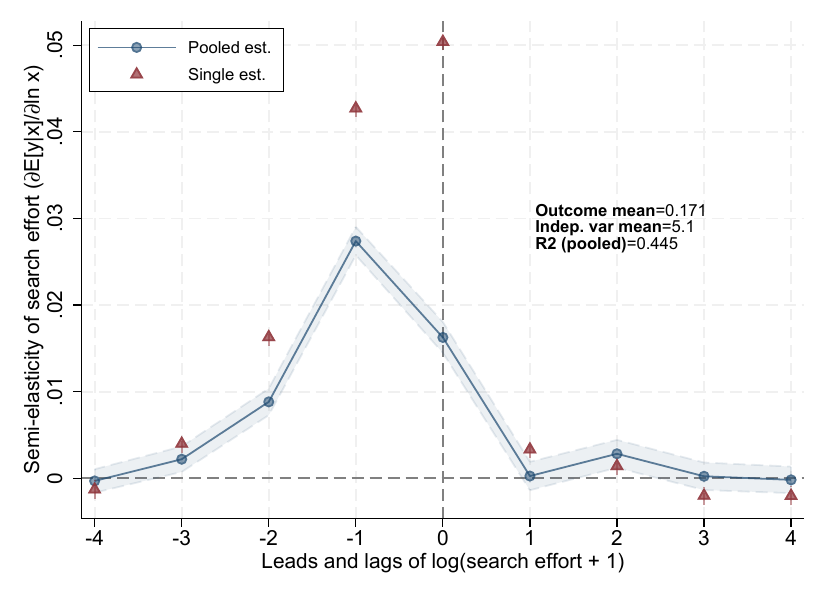}
    \end{subfigure}
    \begin{subfigure}[a]{.49\textwidth}
     \centering
    \caption{Marginal returns}
    \label{fig:searchreturnb}
      \includegraphics[width=\textwidth]{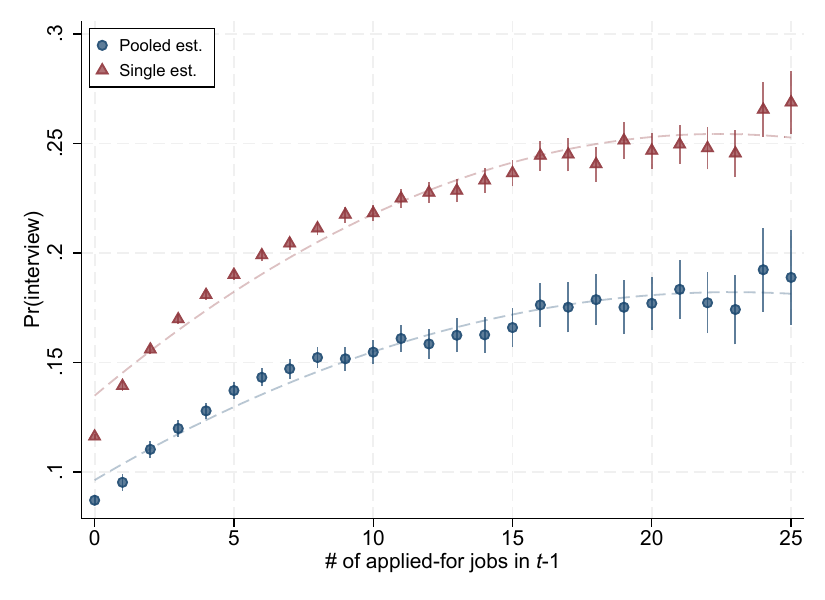}
    \end{subfigure}
    \footnotesize Notes: Panel (a) shows estimates from equation \eqref{eq:searchreturn} of the semi-elasticity of callbacks with respect to search effort by (calendar) months relative to a given month in unemployment. Panel (b) plots the callback probability by number of applied for jobs in the preceding month, residualized for time and spell fixed effects and renormalized to its original mean using the user-written package \texttt{binsreg} \citep[see][]{Cattaneo2025}. The blue dots show estimates when controlling for number of applied-for jobs in surrounding periods wheres the red triangles represents single estimates. Surrounding each estimate are 95\% confidence intervals with standard errors are clustered at the spell level.
\end{figure}

Figure \ref{fig:searchreturna} shows the marginal returns to search effort and their dynamics in the probability of receiving a callback. The blue circles plot estimates of $\phi_{t-k}$, representing the marginal effect of increasing search effort in month $t-k$ while holding effort in surrounding months constant. In contrast, the red triangles show estimates from regressions of the interview attendance indicator on search intensity, run separately for each $k$. These thus capture both the direct impact of applications sent in $t-k$ and the influence of correlated search activity in neighboring periods
In terms of timing, most interviews take place in the calendar month following the submission of applications. Reassuringly, search effort in future periods ($k>0$) is largely uncorrelated with interview attendance in month $t$, suggesting that the fixed-effects design mitigates  dynamic selection. Quantitatively, the estimates imply that increasing search effort by 50\% --- for example, from 4 to 6 applications ---raises the probability of receiving a callback in the following month by roughly 8\%, \textit{ceteris paribus}, with respect to search effort in surrounding periods ($(\frac{0.0274}{100}\times50)/0.171=0.08$). Given that search effort is serially correlated, doubling the number of applications over a three-month period increases the probability of obtaining at least one interview by approximately 30\% ($0.05/0.171 = 0.295$).

Figure \ref{fig:searchreturnb} illustrates the nonlinear relationship between the probability of receiving a callback and the number of applications submitted in period $t-1$ (trimmed at the 99th percentile) using a nonparametric specification. Both variables are residualized with respect to time and spell fixed effects and re-normalized following \cite{Cattaneo2024}. Once again, the blue circles depict the relationship while holding search effort in surrounding months constant, whereas the red triangles also capture the influence of correlated job-search activity in nearby periods. In both cases, the figure indicates diminishing returns to search effort. This pattern supports our functional-form choice in equation \eqref{eq:searchreturn} and aligns with the notion that job seekers initially target the most promising vacancies, while additional applications are directed toward progressively less suitable opportunities.

In ongoing companion work, we investigate the returns to search effort in greater depth and examine heterogeneity across different groups of job seekers \citep{Cederlof2025b}. For the purposes of this paper, however, we focus on showing that job seekers' reported number of applications predicts subsequent callbacks, indicating that this measure captures meaningful variation in their actual search effort.

\section{Duration Dependence} 
\label{sec:duration_dependence}
Having established that search effort remains largely constant within unemployment spells and is predictive of future callbacks, we now turn to how the likelihood of receiving an interview evolves with time in unemployment. This section quantifies the extent to which the observed decline in callbacks reflects “true” duration dependence --- as opposed to dynamic selection --- and outlines a framework for interpreting these findings in relation to duration dependence in job-finding rates.

Following \cite{Mueller2024}, we can decompose observed duration dependence as follows:
    \begin{equation}
    \label{eq:DDdecomp}
        \underbrace{E_d(h_{i,t}) - E_{d+1}(h_{i,t+1})}_{\text{observed duration dependence}} = \underbrace{E_{d}(h_{i,t} - h_{i,t+1})}_{\text{true duration dependence}} + \underbrace{E_{d}(h_{i,t+1}) - E_{d+1}(h_{i,t+1})}_{\text{dynamic selection }}  
    \end{equation} 
where in our case $h_{i,t}$ indicates the probability of individual $i$  attending an interview at time $t$. The first term shows the difference in the expected values between (different samples of) job seekers with unemployment duration $d$ and $d+1$. This can then be decomposed into two parts: The first term right of the equal sign shows the expected change in the probability of an interview for a given sample of individuals between time period $t$ and $t+1$ at duration $d$. This component, often referred to as structural or ``true'' duration dependence, is the direct effect of unemployment duration the probability of getting an interview. The second term, shows the difference in the expected hazard between duration $d$ and $d+1$ which occurs due to compositional changes within the sample. 

Separating these two parts is often a non-trivial endeavor as it requires either estimating the degree of dynamic selection based on job seeker observables \citep[see][]{Mueller2024}, or repeatedly observing $h_{i,t}$ for the same individual within a spell. While duration dependence at its core is often formulated in terms of job-finding (hazard) rates, this per definition cannot be observed more than once per unemployment spell. So, in the spirit of audit studies examining duration dependence in callbacks by randomly varying unemployment duration on fictitious job applicants’ CVs sent to employers \citep{Kroft2013,Eriksson2014,Farber2016}, we exploit job seekers’ reports of attended interviews. For consistency, we refer to these as callbacks throughout. To our knowledge, we alongside \cite{Lalive2025}, are the first to use administrative data on reported interviews to estimate duration dependence in callbacks.

Before turning to the results, it is important to recognize note that although we can identify ``true'' duration dependence in callbacks, how this translates into duration dependence in job-finding rates is less clear. As highlighted by \cite{Jarosch2019}, if time in unemployment carries information about job seeker's unobserved productivity, employers may interpret longer durations as a signal of lower productivity --- knowing that more productive individuals tend to find jobs faster --- and choose not to interview them as the likelihood of a good match is lower. Consequently, many of the interviews withheld from long-term unemployed job seekers would not have materialized into job offers even if they had been granted, thereby weakening the link between observed callback rates and actual job-finding rates.\footnote{It follows from this that, as interviews become more targeted, the job-offer arrival rate --- that is, the number of job offers per attended interview --- may remain constant (or even increase) over time, both at the individual level and in the cross-section. \citet{Lalive2025} provides empirical evidence consistent with this relationship.} In section \ref{sub:connecting_call_backs_and_job_finding} we discuss under what conditions our results on duration dependence in callbacks map onto duration dependence in job-finding rates.

\subsection{Duration dependence in callbacks} 
\label{sub:duration_dependence_in_call_back_rates}

\begin{figure}[t]
  \caption{Pr(interview) by elapsed duration} 
  \label{fig:interview}
    \begin{subfigure}[a]{\textwidth}
     \centering 
      \includegraphics[width=.65\textwidth]{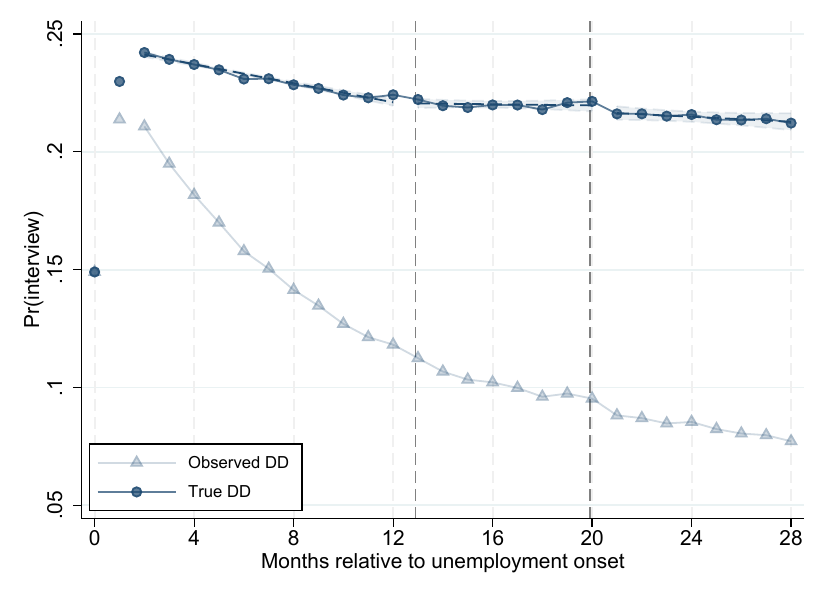}
    \end{subfigure}
\footnotesize Notes: The figure shows the evolution of the probability of attending an interview relative to onset of unemployment. The light colored triangles show estimates from equation \eqref{eq:timeFE} reflecting \textit{observed} duration dependence whereas darker colored circles by solid lines depict estimates from \eqref{eq:spellFE} which accounts for dynamic selection, thus reflecting \textit{true} duration dependence. The colored dashed lines show linear predictions along with 95\% confidence bands, having regressed a dummy for attending an interview on elapsed duration, interacted with an indicator for different segments $k$ of duration $t$ being within the interval $k=\{[2,12],[13,20],[21,28]\}$, including spell fixed effects. All spells are right-censored at the last two calendar months of each unemployment spell. Standard errors are clustered at the spell level.
\end{figure}

Figure \ref{fig:interview} shows how the probability of attending an interview evolves over the course of the unemployment spell. As callbacks evolve similarly (both within and across spells) among job seekers with long or short PBD (see Figure \ref{fig:interviewPBD} in Appendix \ref{app:appendixC}), we have pooled the two groups in our analysis on duration dependence. The light colored triangles show estimates from the benchmark time-fixed effects model (equation \eqref{eq:timeFE}), depicting the average callback rate among the sample of workers remaining in unemployment at month $t$, thus reflecting \textit{observed} duration dependence. As expected, there is a significant decline in the likelihood of getting called for an interview as unemployment continues. At the beginning of the spell ($t+2$), about 21\% of job seekers have attended an interview while in month 12 and 24 of unemployment, the corresponding numbers are 11.8\% and 8.5\%, respectively. 

Naturally, a significant portion of the observed decline reflects compositional changes in the sample of job seekers remaining in unemployment. To separate dynamic selection from the causal effect of unemployment duration on the likelihood of getting an interview (i.e true duration dependence) the darker colored circles in Figure \ref{fig:interview} show estimates including spell-fixed effects (equation \eqref{eq:spellFE}); displaying the within-spell evolution of callbacks. The decline in callbacks is strikingly different, starting at 24.4\% and then declining to merely 22.4\% and 21.6\% one and two years into the spell. These estimates imply on average a 0.77\% or 0.52\% percent decline per month in the likelihood of receiving a callback and also hints at nonlinearities in the duration profile. These results are also very similar to \cite{Eriksson2014} who in an audit study on Swedish employers find a 7.6\% decline in callback rates for workers with 9 months of contemporary unemployment.  

In Table \ref{tab:DDmain}, columns (3) and (4) summarize the estimated callback probabilities from the saturated time-and-spell fixed-effects model. One concern with inferring duration dependence from a fully saturated specification is that the estimates may be sensitive to random month-to-month variation. To address this, columns (1) and (2) of Table \ref{tab:DDmain} report estimates and predicted probabilities from a specification in which unemployment duration enters linearly in equation \eqref{eq:spellFE}, providing a concise summary measure of average duration dependence per month. In panel a) we present estimates on the percentage point change in callback rates based on months 2--12 of unemployment whereas in panel b) the estimates are based on a longer time horizon (months 2--24). In square brackets, we report the implied percentage change in a callback per month, again calculated with compound decay. As shown in column (1), without accounting for dynamic selection, the probability of attending an interview declines by roughly 1 percentage point per month of unemployment, corresponding to a 6.3 percent decrease. When including spell fixed effects, the estimates are nearly an order of magnitude smaller, indicating a decline of about 0.2 percentage points per month, or an average decrease of 0.87 percent. Panel (b) of Table \ref{tab:DDmain} repeats the analysis over a longer horizon of 24 months. This yields smaller average effects, again highlighting the non-linearity of the relationship with most of the duration dependence arising during the first year of the unemployment spell.

To quantify the influence of search effort on duration dependence, we redo the analysis including controls for levels of current and lagged search effort (see Table \ref{tab:DDmainsearch}). This slightly attenuates true duration dependence by about about 15--20\% which is expected given our results in Section \ref{sec:evolution_of_search_behavior} of search effort being predictive of callbacks, but also largely constant over the spell. We interpret this as suggestive evidence that true duration dependence largely originates on the demand side --- that is, from employers statistically discriminating against job seekers with longer unemployment durations.

\begin{table}[t]\centering 
\begin{threeparttable}[b]
\def\sym#1{\ifmmode^{#1}\else\(^{#1}\)\fi} \small
\caption{Estimates of duration dependence} \label{tab:DDmain}
\begin{tabular}{l*{4}{c}}
\toprule
&\multicolumn{2}{c}{Linear model}&\multicolumn{2}{c}{Saturated model} \\
&w/o spell-FE & w. spell-FE &w/o spell-FE & w. spell-FE \\
\cmidrule(lr){2-3} \cmidrule(lr){4-5}  
&(1) & (2) & (3) & (4)  \\
\midrule  
\multicolumn{1}{l}{\textbf{Panel a): Within 12 months}} &&& \\
Elapsed duration&     -0.0098\sym{***}&      -0.0020\sym{***}      &            &            \\
            &    (0.0001)&     (0.0001)       &            &            \\
            &  [-6.310\%]&  [-0.872\%]&  [-5.620\%]&  [-0.767\%]\\
\addlinespace         
  $ \widehat{\tau}_{2}: \widehat{Pr}(interview)_{t+2} $ &       0.204&       0.241&       0.211&       0.242\\
            &     (0.001)&     (0.001)&     (0.001)&     (0.001)\\
  $ \widehat{\tau}_{12}: \widehat{Pr}(interview)_{t+12} $ &       0.106&       0.221&       0.118&       0.224\\
            &     (0.001)&     (0.001)&     (0.001)&     (0.001)\\
\cmidrule(lr){2-3} \cmidrule(lr){4-5}
Share of true dependence&   \multicolumn{2}{c}{13.8\%}&   \multicolumn{2}{c}{13.6\%}\\
\# observations&   \multicolumn{2}{c}{1,341,090}&   \multicolumn{2}{c}{1,341,090}\\
\addlinespace
\multicolumn{1}{l}{\textbf{Panel b): Within 24 months}} &&& \\
Elapsed duration&     -0.0065\sym{***}&    -0.0013\sym{***}        &            &            \\
            &    (0.0001)&     (0.0001)       &            &            \\
            &  [-5.874\%]&  [-0.596\%]&   [4.024\%]&  [-0.522\%]\\
\addlinespace            
  $ \widehat{\tau}_{2}: \widehat{Pr}(interview)_{t+2} $ &       0.193&       0.239&       0.211&       0.242\\
            &     (0.001)&     (0.000)&     (0.001)&     (0.001)\\
  $ \widehat{\tau}_{24}: \widehat{Pr}(interview)_{t+24} $ &       0.051&       0.210&       0.085&       0.216\\
            &     (0.001)&     (0.001)&     (0.002)&     (0.002)\\
\cmidrule(lr){2-3} \cmidrule(lr){4-5}
Share of true dependence&   \multicolumn{2}{c}{10.1\%}&   \multicolumn{2}{c}{13.0\%}\\
\# observations&   \multicolumn{2}{c}{1,724,838}&   \multicolumn{2}{c}{1,724,838}\\
\bottomrule
\end{tabular}
\begin{tablenotes}[flushleft]
\scriptsize \item \noindent  Notes: The table shows how the probability of attending an interview evolves during within 12 (panel a) and 24 (panel b) months of unemployment. Columns (1) and (2) show estimates from regressing a dummy of having a attended an interview on a continuous variable of elapsed unemployment duration, with and without spell-fixed effects, respectively. Columns (3) and (4) show $\tau$-estimates from equation \eqref{eq:timeFE} and \eqref{eq:spellFE}, respectively. In hard brackets are the implied average percentage change per month calculated with compound decay formula: $
\Delta_k\% = 1 - \left(\hat{y}_{\overline{k}}/\hat{y}_{\underline{k}} \right)^{\frac{1}{\overline{k} - \underline{k}}}$ where $k \in \{[2,12], [2,24]\}$. All spells are right-censored at the last two calendar months of each unemployment spell. Standard errors are in parenthesis and clustered at the spell level. Asterisks indicate that the estimates are significantly different from zero at the  \sym{*} \(p<0.1\), \sym{**} \(p<0.05\), \sym{***}  \(p<0.01\) level.
\end{tablenotes}
\end{threeparttable}
\end{table}

We have investigated the robustness of our results in multiple ways. First, we evaluate the modeling assumption of no dynamic selection in changes by plotting the probability of getting called for an interview separately for groups of job seekers with different realized unemployment durations. As shown in Figure \ref{fig:interviewfixed} in the Appendix \ref{app:appendixC}, although these groups differ in the level of likelihood of being called to an interview, they evolve very much in parallel over the course of the spell, suggesting that the spell-fixed effect model is able to handle most of the dynamic selection occurring over time. Second, we have estimated duration dependence with and without spell-fixed effects using Poisson regression which show similar results as the OLS (see Figure \ref{fig:poisson} in Appendix \ref{app:appendixC}). Third, we have used the number of attended interviews in a given month as our dependent variable which yields virtually identical results as the bulk of job seeker either have zero or one interview in a given month. 

Our results suggest that true duration dependence plays only a limited role in explaining the observed decline in callbacks. Depending on the time horizon, 86--90\% of this decline can be accounted for by compositional changes among job seekers (i.e., dynamic selection), leaving only 10--14\% attributable to true duration dependence. These magnitudes closely mirror those reported by \cite{Mueller2024} for duration dependence in \textit{job-finding rates}. They find that 85\% of the observed decline among Swedish job seekers reflects dynamic selection. While \cite{Mueller2024} infers true duration dependence indirectly --- by estimating the degree of dynamic selection --- our approach identifies it directly. The close alignment of the two sets of estimates, despite their methodological differences and distinct outcomes, reinforces the conclusion that dynamic selection accounts for the bulk of the decline in job-search success over the unemployment spell. However, our findings contrast recent evidence by \cite{Lalive2025} who find that only about a third of observed duration dependence in callbacks can be attributed to true duration dependence whereas for job-finding rates, the two forces contribute in almost equal proportions. It is of course difficult to exactly pin down why our results and those of \cite{Mueller2024} results differ from those of \cite{Lalive2025}. In light of our results on heterogeneity below, one possible candidate is differences in labor market tightness. Whereas the average unemployment rate in Switzerland during the sample period in \cite{Lalive2025} was 4.73\%, during our sample period the unemployment rate averaged around 7.3\%. Another key component may be differences in search effort which appear to be decreasing over the unemployment spell among Swiss job seekers.\footnote{Even though \cite{Lalive2025} do not censor final parts of the unemployment spells in their sample -- which we show to be an important explanation for observing a declining search effort within the spell -- they show (in the Appendix) that the number of applications is indeed decreasing within the unemployment spell by plotting search effort over realized unemployment durations. In their analysis of job-finding they attribute about 85\% of the decline to reduced search effort among job seekers.} A final reason may be that our analysis, using spell-fixed effects, is more effective in controlling for compositional changes arising over time.

\subsection{Connecting degree of duration dependence in callbacks and job-finding rates} 
\label{sub:connecting_call_backs_and_job_finding}
The coherence between our results and those of \cite{Mueller2024} suggests that the \textit{share} of true duration dependence in callbacks may serve as an approximation of the \textit{share} of true duration dependence in job-finding rates, despite the potentially weak direct link between callbacks and actual job-finding outcomes \citep{Jarosch2019}. To formalize this idea, we introduce a simple stylized framework of job-finding and define the job-finding rate ($h$) for a given job seeker as:
\begin{equation}
\label{eq:indhaz}
	h(x,t) = \lambda_o(x,t)\lambda_i(x,t)
\end{equation}
where $\lambda_o$ is the job-offer arrival rate conditional on getting an interview and $\lambda_i(x,t)$ is the probability of being called to an interview, both being a function of elapsed time in unemployment $t$ and job seeker characteristics $x\in \mathcal{X}$ at time $t$ with a cross-sectional density function $f(x;t)$ normalized such that $\int_{\mathcal X} f(x;t)\,dx=1$.\footnote{Given our results on search effort being largely constant both across and within a spell, we abstain from modelling the callback probability  as a function search effort explicitly.} The average job-finding rate in the cross-section at time $t$ is then given by $\bar{h}(t)=\int_{\mathcal X} h(x,t)\, f(x;t)\,dx$, whereas the average callback probability and job-finding rate in the cross-section can be written as $\bar\lambda_k(t)=\int_{\mathcal X}\lambda_j(x,t)\,f(x;t)\,dx$ with $j=\{i,o\}$. For simplicity, we follow \cite{Jarosch2019} and assume that no job offers arrive unless having been preceded by an interview and once a job offer arrives it is accepted by the job seeker.\footnote{Assuming that job seekers accept all job offers seems plausible in our setting given that they  face UI suspension if they decline a pending job offer.} The derivative of equation \eqref{eq:indhaz} defines true duration dependence and its share relative to observed duration dependence can then be written as: 
\begin{equation}
	\label{eq:shareDD}
	\mathcal R_h(x,t) \equiv
\frac{\partial_t h(x,t)/\partial t}{\,d\bar h(t)/dt\,}
=
\frac{(\partial_t\lambda_o)\lambda_i+\lambda_o(\partial_t\lambda_i)}
{\displaystyle \int_{\mathcal X}\!\Big[(\partial_t\lambda_o)\lambda_i f
+ \lambda_o(\partial_t\lambda_i) f
+ \lambda_o\lambda_i(\partial_t f)\Big]\,dx}
\end{equation}
where $\partial_t$ denotes the partial derivative with respect to $t$. As can be seen in equation \eqref{eq:shareDD}, it contains the share of true duration dependence in callbacks:
\begin{equation}
\mathcal R_i(x,t) \equiv	\frac{\partial\lambda_i(x,t)/\partial t}{\,d\bar\lambda_i(t)/dt\,}
=
\frac{\partial_t \lambda_i(x,t)}
{\displaystyle \int_{\mathcal X}\!\Big[(\partial_t\lambda_i) f
+ \lambda_i(\partial_t f)\Big]\,dx},
\end{equation}
Again, our empirical findings suggest that $\mathcal{R}_h(x,t) \approx \mathcal{R}_i(x,t)$, which is consistent with our framework under two conditions. First, the job-offer arrival rate must be approximately constant over duration (i.e., $\partial \lambda_o / \partial t \approx 0$). Alternatively, it suffices that this derivative is small relative to the change in the callback probability, both at the individual level and in the cross-section. Second, there must be limited heterogeneity in job-offer arrival rates at any given duration. We show formally in Appendix~\ref{app:appendixD} that, under these conditions, $\mathcal{R}_h(x,t) \approx \mathcal{R}_i(x,t)$. Intuitively, the first condition implies that $\partial_t \lambda_o$ is negligible and can thus be ignored, whereas the second condition implies that the ratio of job-offer arrival rates in the numerator and denominator of equation~\eqref{eq:shareDD} cancels out.

While we cannot empirically verify these conjectures ourselves as we do not observe job-offers in our data, we note that both of these assumptions are consistent with the mechanisms in \cite{Jarosch2019} where employers statistically discriminate towards the long-term unemployed, basing their callback decisions on job seekers' duration in unemployment. Moreover, the empirical evidence in \cite{Lalive2025} suggests that job-offer arrival rates are indeed largely constant over the duration of unemployment, both within and across spells.

\section{Heterogeneity in search effort and callbacks} 
\label{sec:heterogeneity_in_search_and_call_back_rates}

We now turn to exploring heterogeneity in search effort and callback rates across different groups of job seekers and labor market states, and in particular, in its evolution over the unemployment spell. In doing so, we continue to restrict our attention to the first 12 months of unemployment, pooling job seekers with different PBD eligibility. Doing so has several advantages: $i)$ it avoids conflating our estimates with the observed drop in search effort around UI exhaustion and differences in PBD eligibility across groups\footnote{We have looked at differences in search intensity at UI exhaustion between groups and find them to be largely proportional (results are available upon request).}; $ii)$ it focuses on the part of the unemployment spell where the majority of duration dependence in callbacks manifests; and $iii)$ simplifies comparison to existing audit studies which study differences in callback rates during the first year in unemployment \cite[see e.g.][]{Kroft2013,Eriksson2014}.

Table \ref{tab:hetro} presents the within-spell linear predictions of number of applied-for jobs (panel a) and probability of an interview (panel b) along with the implied percentage changes per month.\footnote{Using a fully saturated model we arrive at very similar conclusions. Figure \ref{fig:hetrosearch} and Figure \ref{fig:hetrodur} in Appendix \ref{app:appendixB} show the graphical evidence of our heterogeneity analysis from both the linear and saturated model for search effort and callback rates, respectively} Columns (1) and (2) show results for native and non-native job seekers, respectively. At baseline, natives apply for 21\% fewer jobs than to non-natives, amounting to roughly 1.5 jobs per month. During the first 12 months of unemployment natives lower their search intensity by on average 0.56\% per month whereas non-natives exhibits a slight increase over time by 0.15\%. Interestingly, we observe the reverse pattern in callbacks. Here natives start out at a callback rate of 0.27 and non-natives just below 0.19, implying a 30\% callback gap. Importantly, while our results are similar to the racial or ethnic callback gaps found in audit studies, we want to stress that our estimates cannot serve as a standalone indicator of employer discrimination as other confounders may contribute to the observed gap (in either direction).\footnote{For the US, \cite{Bertrand2004} find that black-sounding names received about 33\% fewer callbacks than white-sounding names. Audit studies performed in Sweden, studying differences in callback rates between Swedish and Arabic-sounding names find a callback gap of about 50\% \citep{Carlsson2007,Adermon2022}.} Both natives and non-natives experience duration dependence with callback rates decreasing by 0.81\% for natives and 0.95\% for non-natives, on a monthly basis. Relative to baseline, 12 months into unemployment the callback rate has in total declined by 7.7\% for natives and 9\% for non-natives.\footnote{Splitting up non-natives into EU and non-EU immigrants, we find that the former have on average a 78\% higher callback probability compared to non-EU immigrants (27.3\% vis-a-vis 15.3\%) at the same levels as natives. However, we find no differences in the degree of experienced duration dependence between EU and non-EU immigrants.} Taken at face value, this implies that non-natives experience about 17\% higher degree of duration dependence. 

Columns (3) and (4) show estimates separately for male and female job seekers. We note that while females apply for roughly 0.8 additional jobs per month, there is no statistically significant difference in the month-to-month change. In terms of callback rates, both male and female job seekers lower their probability of getting called to an interview by 0.002 percentage points per month. Accounting for the initial level differences between the two groups this corresponds to a monthly decrease by 0.96\% for males and 0.77\% for females.

\begin{landscape}
\begin{table}[p]\centering 
\begin{threeparttable}[b]
\def\sym#1{\ifmmode^{#1}\else\(^{#1}\)\fi} \small
\caption{Heterogeneity in search effort and duration dependence (linear model)} \label{tab:hetro}
\begin{tabular}{l*{8}{c}}
\toprule
&\multicolumn{2}{c}{Nativity status}&\multicolumn{2}{c}{Gender}&\multicolumn{2}{c}{Target occupation}&\multicolumn{2}{c}{LM tightness} \\
&Native & Non-native & Male & Female &High-skilled & Low-skilled & Tight & Slack \\
\cmidrule(lr){2-3} \cmidrule(lr){4-5} \cmidrule(lr){6-7} \cmidrule(lr){8-9} 
&(1) & (2) & (3) & (4) &(5)& (6)& (7) &(8) \\
\midrule 
\multicolumn{2}{l}{\textbf{Panel a): Search effort}} &&&&&& \\
\addlinespace
Elapsed duration&     -0.0281\sym{***}&      0.0101\sym{**} &     -0.0078\sym{***}&     -0.0058         &     -0.0189\sym{***}&     -0.0007         &     -0.0018         &     -0.0120\sym{***}\\
            &    (0.0029)         &    (0.0040)         &    (0.0030)         &    (0.0041)         &    (0.0040)         &    (0.0031)         &    (0.0036)         &    (0.0030)         \\
            &  [-0.560\%]         &   [0.153\%]         &  [-0.147\%]         &  [-0.096\%]         &  [-0.312\%]         &  [-0.013\%]         &  [-0.029\%]         &  [-0.268\%]         \\
            \addlinespace
 $ \text{applied-for jobs}_{t+2} $ &       5.146         &       6.525         &       5.337         &       6.103         &       6.140         &       5.322         &       6.254         &       4.549         \\
 $ \text{applied-for jobs}_{t+12} $ &       4.865         &       6.626         &       5.259         &       6.045         &       5.951         &       5.315         &       6.237         &       4.428         \\
 \addlinespace \addlinespace
\multicolumn{2}{l}{\textbf{Panel b): Duration dependence}} &&&&&& \\
Elapsed duration&     -0.0021\sym{***}&     -0.0017\sym{***}&     -0.0020\sym{***}&     -0.0020\sym{***}&     -0.0023\sym{***}&     -0.0018\sym{***}&     -0.0025\sym{***}&     -0.0014\sym{***}\\
            &    (0.0002)         &    (0.0002)         &    (0.0001)         &    (0.0002)         &    (0.0002)         &    (0.0001)         &    (0.0002)         &    (0.0002)         \\
            &  [-0.810\%]         &  [-0.955\%]         &  [-0.956\%]         &  [-0.773\%]         &  [-0.779\%]         &  [-0.956\%]         &  [-1.007\%]         &  [-0.657\%]         \\
            \addlinespace
  $ \widehat{Pr}(interview)_{t+2} $ &       0.273         &       0.187         &       0.221         &       0.268         &       0.303         &       0.196         &       0.255         &       0.214         \\
  $ \widehat{Pr}(interview)_{t+12} $ &       0.252         &       0.170         &       0.201         &       0.248         &       0.280         &       0.178         &       0.231         &       0.200         \\
\midrule \# observations & 749,331 & 589,303 & 793,469 & 547,621 & 518,852 & 822,238 & 823,851 & 517,239 \\
\bottomrule
\end{tabular}
\begin{tablenotes}[flushleft]
\scriptsize \item \noindent  Notes: The table show heterogeneity in within-spell estimates on the evolution of number of applied-for jobs (panel a) and the probability of attending an interview (panel b), within the first year of unemployment. These come from regressing the outcome on a continuous variable of elapsed unemployment duration interacted with a dummy variable for the duration segment month 2--12, separately for each group of job seekers indicated in the column headings. In hard brackets we show the average month-to-month percent change in the outcome which we calculate by predicting the outcomes at $t+2$ and $t+12$ and use the compound decay formula: $\Delta_k\% = 1 - \left(\hat{y}_{\overline{k}}/\hat{y}_{\underline{k}} \right)^{\frac{1}{\overline{k} - \underline{k}}}$ where $k \in\{2,12\}$. All spells are right-censored at the last two calendar months of each unemployment spell. Standard errors are in parenthesis and clustered at the spell level. Asterisks indicate that the estimates are significantly different from zero at the  \sym{*} \(p<0.1\), \sym{**} \(p<0.05\), \sym{***}  \(p<0.01\) level.
\end{tablenotes}
\end{threeparttable}
\end{table}

\end{landscape}

Next, we use data on job seekers preferred occupations which they state when registering at the PES and split job seekers by whether they are looking for a high or low-skilled job.\footnote{A job seeker may state up to four different preferred occupations when registering at the PES. These are in the format of four digit occupational codes (Swedish equivalent of the ISCO codes, SSYK) where the first digit represents the level of qualification needed for the job. We define a job seeker as looking for a high-skilled occupation if at least one out of the potential four preferred occupations requires higher level education. The share of job seekers stating at least one high-skilled occupation as one of their preferred occupations is 28\%.} Column (5) and (6) display the estimates showing that, at baseline, job seekers looking for high-skilled jobs apply for almost for one more job per month compared job seekers looking for more low-skilled jobs (corresponding to 15\% higher search intensity). However, while job seekers aiming for low-skilled jobs uphold a constant search effort over the first 12 months of unemployment, job seekers targeting high-skilled occupations lower their search intensity by -0.31\% per month. In terms of differences in callbacks, job seekers targeting high-skilled occupations have roughly twice as high probability of being called for an interview at the beginning of the spell compared to job seekers targeting low-skilled occupations. Their duration dependence is also -18.5\% lower than that of those targeting low-skilled occupations. One possible interpretation of this is that employers places less weight previous unemployment duration when recruiting for a high-skilled position, being aware of that matching between employers and job seekers for such positions takes longer time and thus unemployment duration being less of a negative signal. Interestingly, the probability of being called for an interview continues to decrease among job seekers targeting high skilled jobs also after 12 months of unemployment, whereas for job seekers targeting low-skilled jobs flattens out (see Figure \ref{fig:hetrodur} in Appendix \ref{app:appendixC}). 

Finally, columns (7) -- (8) show the evolution of search effort and callback rates by labor market tightness. Using regional data on quarterly vacancy rates, we define unemployment spells as facing a slack (tight) labor market if the average vacancy rate during the spell is below (above) the median in the distribution within the region. Job seekers looking for jobs in a tight labor market apply for more jobs and uphold the same search effort throughout the spell. Meanwhile, we observe a small, but significant, decrease in the number of applied-for jobs during the first year in unemployment among job seekers looking for work in a slack labor market. In panel b) we see that, as expected, the average callback rate is around 19\% higher in tight labor markets. Interestingly, the likelihood of getting a callback in a given month have during the course of a year decreased by 9.4\% among job seekers looking for jobs in a tight labor market, whereas the corresponding number for job seekers operating in a slack labor market is 6.5\%. This means that duration dependence is just about 50\% higher in tight labor markets. These findings echoes those of \cite{Kroft2013} showing that similar job applications with different unemployment durations randomly assigned on their CV:s, experience less duration dependence when the unemployment rate is high and labor markets are slack. A likely explanation for these results is that an employer finds the length of the job seeker's unemployment spell to be weaker signal of productivity when knowing the worker has faced a weak labor market. 

Overall, when it comes to the evolution of search effort, even though the aforementioned differences are often statistically significant, they appear quite small and of little economic importance. On the other hand, effects on callbacks (i.e duration dependence) are larger with declines of up to 1\% per month. Again, the economic consequences of these declines will nevertheless depend on the what share of interviews translate into a job offer. We devote the next subsection to explore heterogeneity in age; where we see a clear gradient both in search effort and duration dependence.

\subsection{Age gradients in search effort and callbacks} 
\label{sub:age_gradients_in_search_and_call_backs}
In recent years, several OECD countries have experienced persistently high youth unemployment alongside growing concerns about barriers faced by older job seekers in re-entering employment, often linked to age-related biases and “ageism” \citep{OECD2025}.  These contrasting challenges highlight the importance of examining how job search behavior and duration dependence callbacks vary across the age distribution.

\begin{figure}[t]
  \caption{Evolution of search effort by age} 
  \label{fig:searchage}
    \begin{subfigure}[a]{.49\textwidth}
     \centering 
    \caption{Young vs. old}
     \label{fig:searchagea}
      \includegraphics[width=\textwidth]{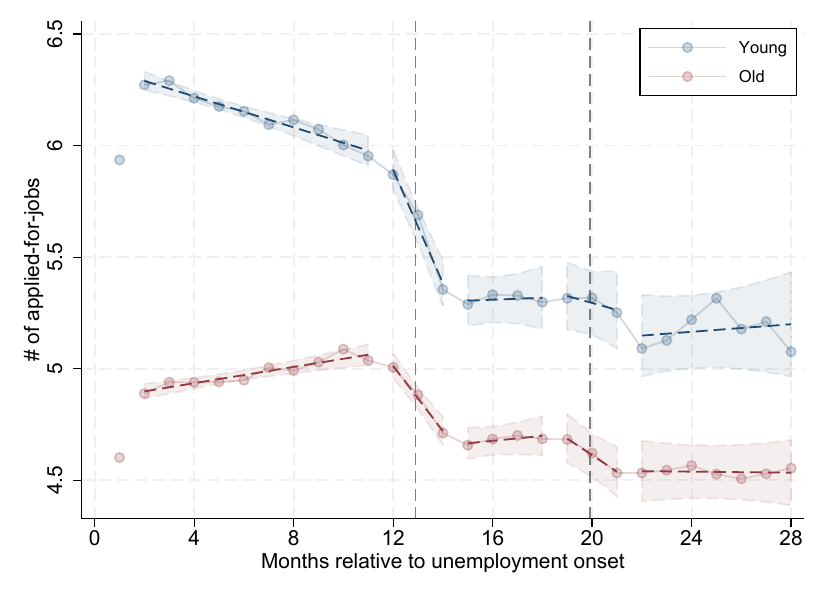}
    \end{subfigure}
    \begin{subfigure}[a]{.49\textwidth}
     \centering
    \caption{Average monthly percentage change}
    \label{fig:searchageb}
      \includegraphics[width=\textwidth]{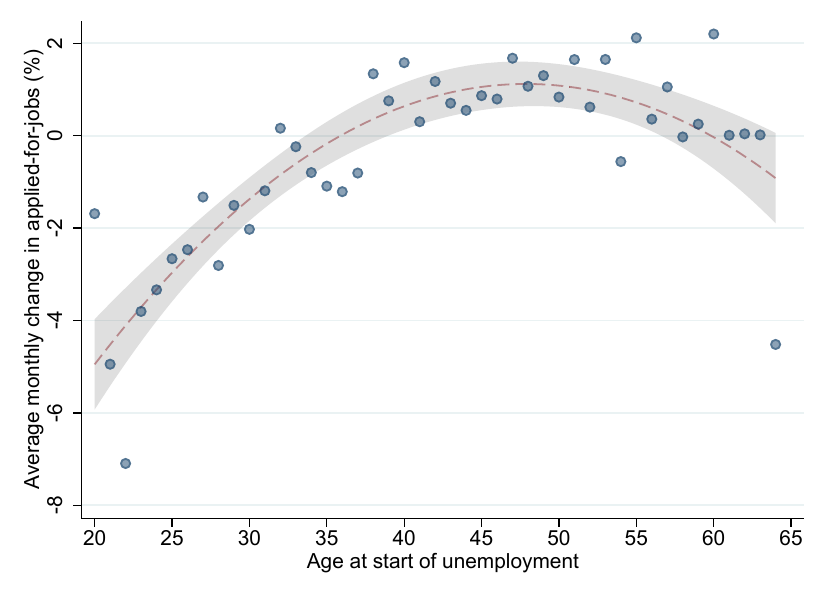}
    \end{subfigure}
\footnotesize Notes: Panel (a) shows within spell variation in the number of applied-for jobs by unemployment duration. The blue and red dots show estimates from equation \ref{eq:spellFE} run separately for young and old job seekers, respectively, defined as being below or above the median in the sample (39). The colored dashed lines show linear predictions along with 95\% confidence bands, having regressed number of applied-for jobs on elapsed duration, interacted with an indicator for different segments $k$ of duration $t$ being within the interval $k=\{[2,11],[12,14],[15,18],[19,21],[22,28]\}$. Standard errors are clustered at the spell level. Panel (b) shows the within-spell average month-to-month percent change, in number of applied-for jobs during the first year of unemployment separately estimated for each age cohort. Estimates come from regressing number of applied-for jobs on a continuous variable of elapsed unemployment duration and then taking the linear predictions to calculate the implied average percentage change per month using the compound decay formula: $\Delta_k\% = 1 - \left(\hat{y}_{\overline{k}}/\hat{y}_{\underline{k}} \right)^{\frac{1}{\overline{k} - \underline{k}}}$ where $k \in\{2,12\}$. The dashed red line show a uniformly weighted quadratic fit over the estimates along with a 95\% prediction interval. For visual purposes panel (a) contain only job seekers eligible to 14 months of PBD whereas panel (b) uses the full sample. All regressions control for time and spell fixed effects and spells are right-censored at the last two calendar months of each unemployment spell.
\end{figure}

We start by depicting the evolution of search intensity by age in Figure \ref{fig:searchagea}, using a simple division between young and old job seekers, split at the median of the distribution at age 39. For visual purposes, this figure restricts attention to job seekers eligible to 14 months of PBD. Young job seekers start out initially applying for about 30\% more jobs than older job seekers (6.3 vs. 4.9 per month) and their search intensity declines by about 0.56\% per month during the first year of unemployment. In contrast, older job seekers increase their intensity by around 0.37\% per month. However, this coarse division into young and old masks a great deal of heterogeneity. Figure \ref{fig:searchageb}  instead plots the within-spell average month-to-month percent change, in number of applied-for jobs during the first year of unemployment over the entire age distribution (for the full sample). This reveals a clear increasing and concave relationship between the change in search effort and age. While the youngest job seekers lower their search effort by on average -5\% per month, search effort increases by age, turns positive and flattens out around age 38.

  \begin{figure}[t!]
    \caption{Callbacks and duration dependence by age}
  \label{fig:DDhetroage}
    \begin{subfigure}[a]{.49\textwidth}
     \centering
     \caption{Pr(interview) (raw)}
     \label{fig:DDhetroagea}
      \includegraphics[width=\textwidth]{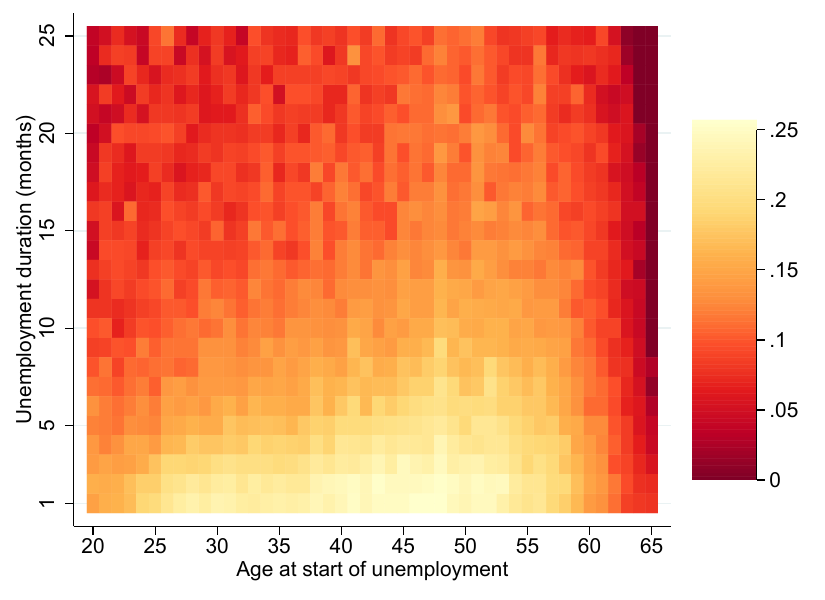}
        \end{subfigure}
        \begin{subfigure}[a]{.49\textwidth}
         \centering
      \caption{Pr(interview) $\mid \alpha_i$}
      \label{fig:DDhetroageab}
      \includegraphics[width=\textwidth]{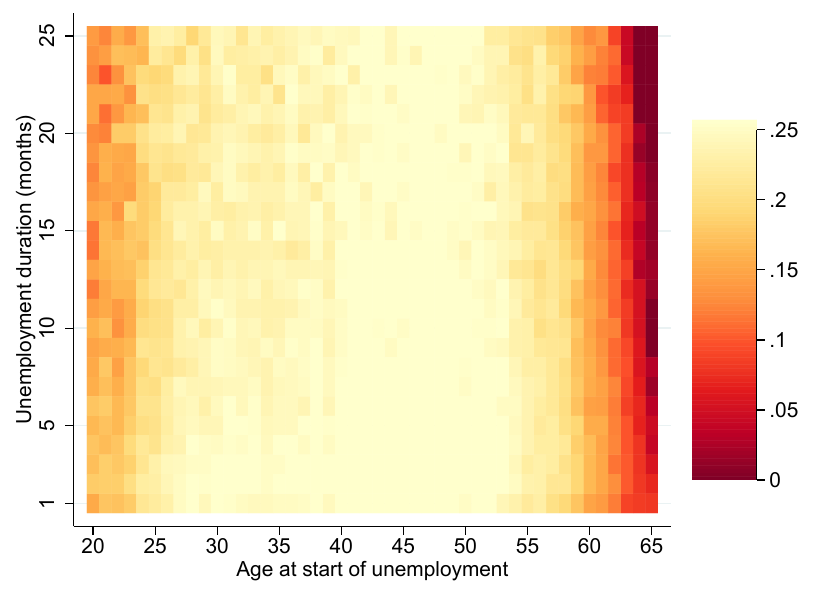}
      \end{subfigure}
     \begin{subfigure}[a]{\textwidth}
         \centering
      \caption{`True' duration dependence}
      \label{fig:DDhetroageac}
      \includegraphics[width=.5\textwidth]{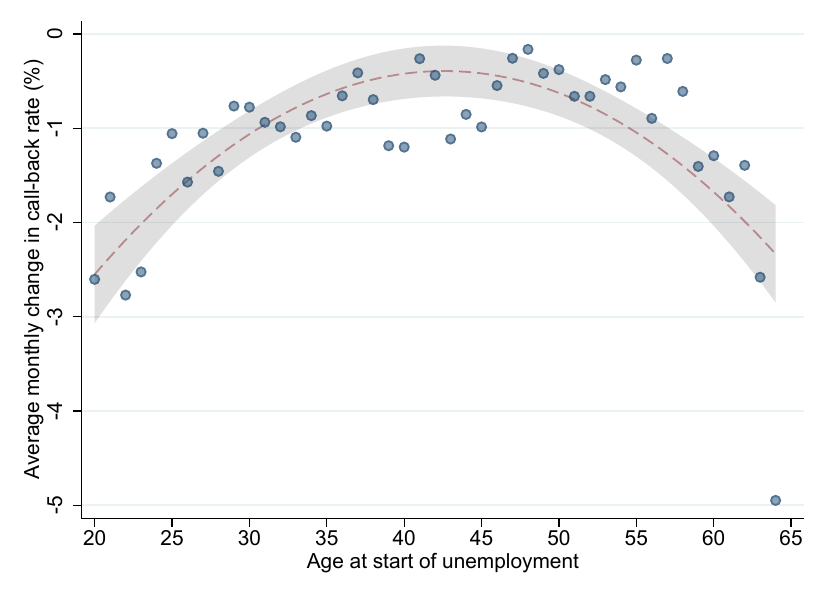}
      \end{subfigure}
   \scriptsize Notes: Panel (a) plots the probability of having attended an interview by month of elapsed duration and age with the coloring of red to yellow going from low to high probability, respectively. Panel (b) plots the same probability while holding constant spell-fixed effects. Panel (c) shows the within-spell average month-to-month percent change, in the probability of attending an interview during the first year of unemployment separately estimated for each age cohort. These come from regressing a dummy of having a attended an interview on a continuous variable of elapsed unemployment duration, including time and spell fixed effects. We then take the linear predictions of this model and calculate the implied average percentage change per month using the compound decay formula: $\Delta_k\% = 1 - \left(\hat{y}_{\overline{k}}/\hat{y}_{\underline{k}} \right)^{\frac{1}{\overline{k} - \underline{k}}}$ where $k \in\{2,12\}$. The dashed red line show a uniformly weighted quadratic fit over the estimates along with a 95\% prediction interval. All spells are right-censored at the last two calendar months of each unemployment spell.
  \end{figure}

Next, we turn to duration dependence in callbacks where earlier correspondence studies have found significant statistical discrimination by the age of the job seeker \citep[see e.g.][]{Lahey2008,Neumark2019,Carlsson2019}. Figure \ref{fig:DDhetroagea} plots in a heat-map raw callback probabilities by age and elapsed duration, showing that job seekers at longer durations have significantly lower likelihood of having been called for an interview. This, however, blends together true duration dependence and dynamic selection and once conditioning on spell-fixed effects almost the entire decline in callbacks vanishes for middle-aged job seekers (see Figure \ref{fig:DDhetroageab}). However, among individuals near retirement, the pattern remains virtually identical to that of without spell-fixed effects and among the young a smaller negative gradient is visible. Figure \ref{fig:DDhetroageac} plots the average month-to-month percent change in callback probabilities by job seekers' age. While the youngest job seekers experience around a 2\% monthly decline in callbacks, the corresponding number is -0.6\% among middle-aged workers. The degree of duration dependence again starts to decline at age 59 and amounts to about -2.2\% per month. 

To what extent can differences in true duration dependence be accounted for by differences in search effort? We investigate this by estimating duration dependence controlling for age-specific trends (and levels) of search effort. Figure \ref{fig:DDhetroage_cont_search} in Appendix \ref{app:appendixC} plots the results. Interestingly, while controlling for search effort lowers true duration dependence about by 6\% on average, the decrease is more than twice as big (14\%) among the youngest job seekers (aged 20--25). So, while some of the observed difference in duration dependence across ages can be attributed to differences in search effort, most of it appears to stem from others factors, such as e.g. statistical discrimination occurring due to ``ageism''.

\section{Concluding remarks} 
\label{sec:conclusions}
Understanding how job seekers adjust their search behavior over the course of unemployment, and the extent to which duration dependence shapes the hazard of reemployment, is central to models of labor market dynamics and the design of public policy. Despite this importance, our understanding of these mechanisms remains limited. This paper leverages rich administrative data on 2.4 million monthly activity reports from Swedish unemployment benefit recipients to provide new evidence on how search behavior and callbacks evolve during unemployment. By exploiting the repeated nature of the data using a time-and-spell fixed effects design, we are able to separate between within-spell changes and dynamic selection.

Three main results emerge. First, once dynamic selection is accounted for, job seekers’ search effort is largely constant throughout unemployment, with a sharp decline only in the two months preceding reemployment. Around UI exhaustion, search intensity drops by about ten percent, likely because participation in active labor market programs crowds out job search. Second, reported interviews confirm that job seekers’ search activity is meaningful: higher search effort predicts higher callback probabilities, showing clear and diminishing returns to search. Third, while raw callback probabilities decline steeply with unemployment duration, most of this decline—about 85 to 90 percent—reflects dynamic selection, leaving a relatively modest role for “true” duration dependence. Most of it appear to be demand-driven, with employers statistically discriminating long-term unemployed job seekers. Under plausible conditions, we also show that the share of true duration dependence in callbacks is informative of that in job-finding rates, suggesting that the causal impact of elapsed unemployment on reemployment hazards is limited.

Finally, we document marked heterogeneity in both search behavior and duration dependence. Search effort varies systematically with age, declining steeply among younger workers but remaining stable or even increasing among older job seekers, while callback probabilities exhibit an inverse U-shaped relationship with age. Moreover, job seekers in tight labor markets experience about 50 percent stronger duration dependence than those in slack markets.

Taken together, our findings suggests that most of the observed decline in job-finding rates with unemployment duration reflects dynamic selection rather than adjustments to search or duration dependence. One limitation of our analysis is that we cannot observe changes in job search behavior unrelated to the number of applications but that influence job seekers probability of reemployment; such as which types of jobs job seekers apply for or whether they lower their reservation wages over the spell. A promising avenue for future research would be to link administrative data on job search with vacancy-level information to better understand how the content of search behavior and employer screening jointly shape reemployment prospects over the unemployment spell.

\clearpage
\newpage

\bibliographystyle{chicago}
\bibliography{ref2}
\newpage 

\addcontentsline{toc}{section}{Online Appendix}
\appendix
\section{Additional descriptives} 
\label{app:appendixA}
\counterwithin{figure}{section}
\counterwithin{table}{section}
\counterwithin{equation}{section}
\setcounter{table}{0}
\setcounter{figure}{0}

\begin{figure}[htbp]
	\caption{Example of activity report}
	\label{fig:actreport}
	    \begin{subfigure}[a]{\textwidth}
	    \centering
      	\includegraphics[page=1, width=0.9\textwidth]{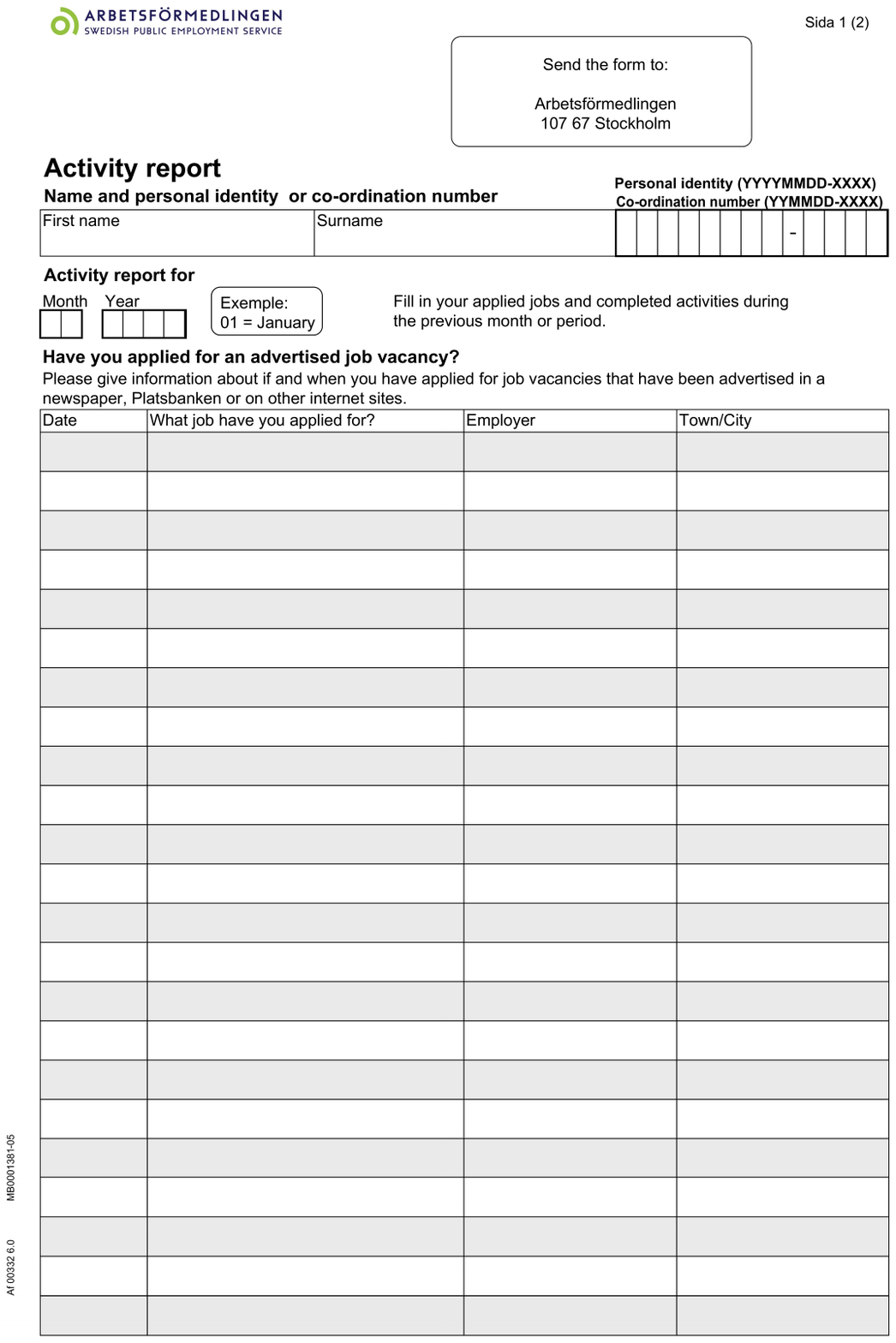}
    \end{subfigure}
\end{figure}
\begin{figure}[htbp]
	    \begin{subfigure}[a]{\textwidth}
	    \centering
      	\includegraphics[page=2, width=0.9\textwidth]{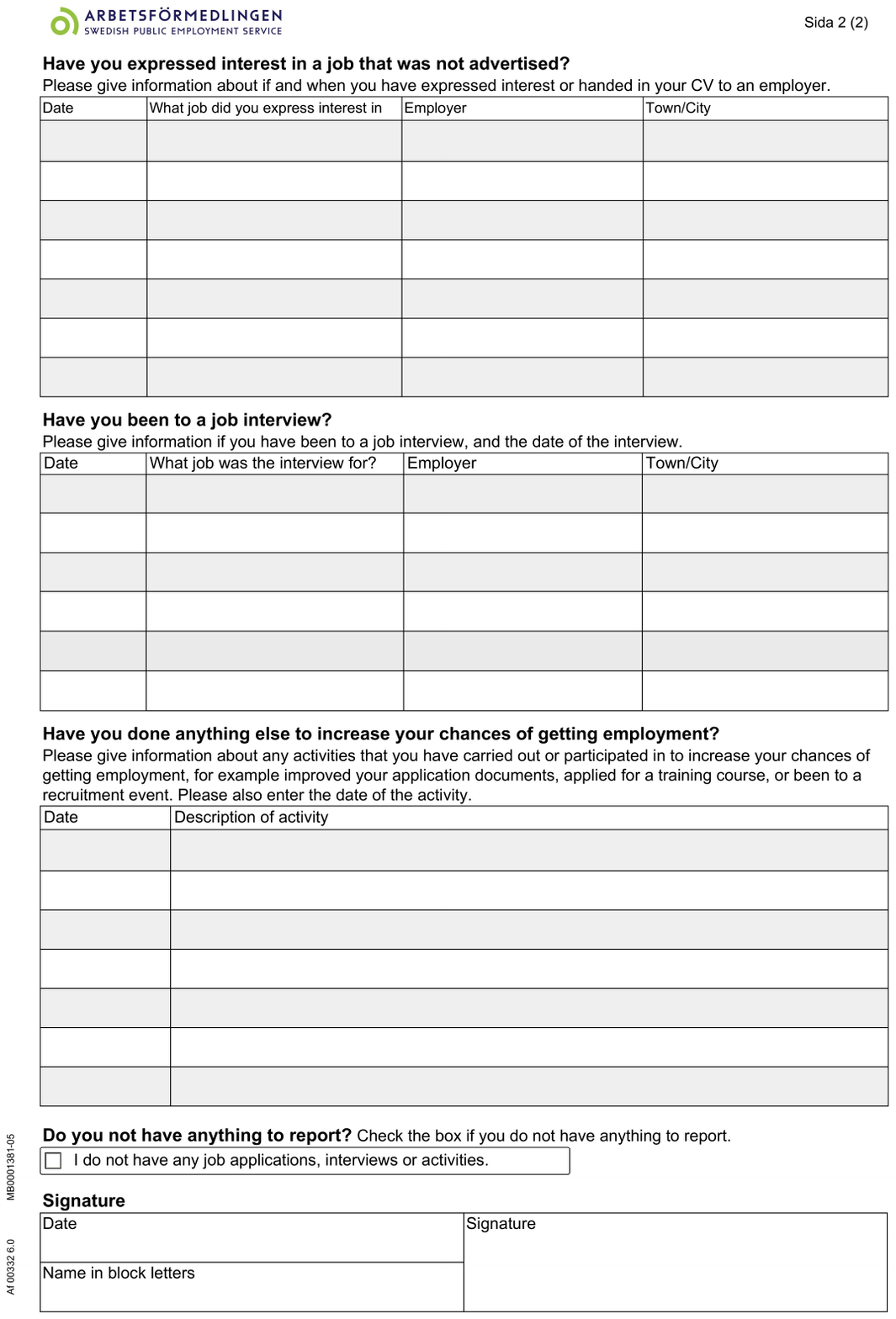}
    \end{subfigure}
    \footnotesize Notes: The figure shows an example of an activity report.
\end{figure}
\newpage
\setcounter{figure}{1}

\begin{figure}[htbp]
	\caption{Example of online activity reporting}
	\label{fig:actreportonline}
		   \begin{subfigure}[a]{\textwidth}
	    \centering
     	\includegraphics[width=0.85\textwidth]{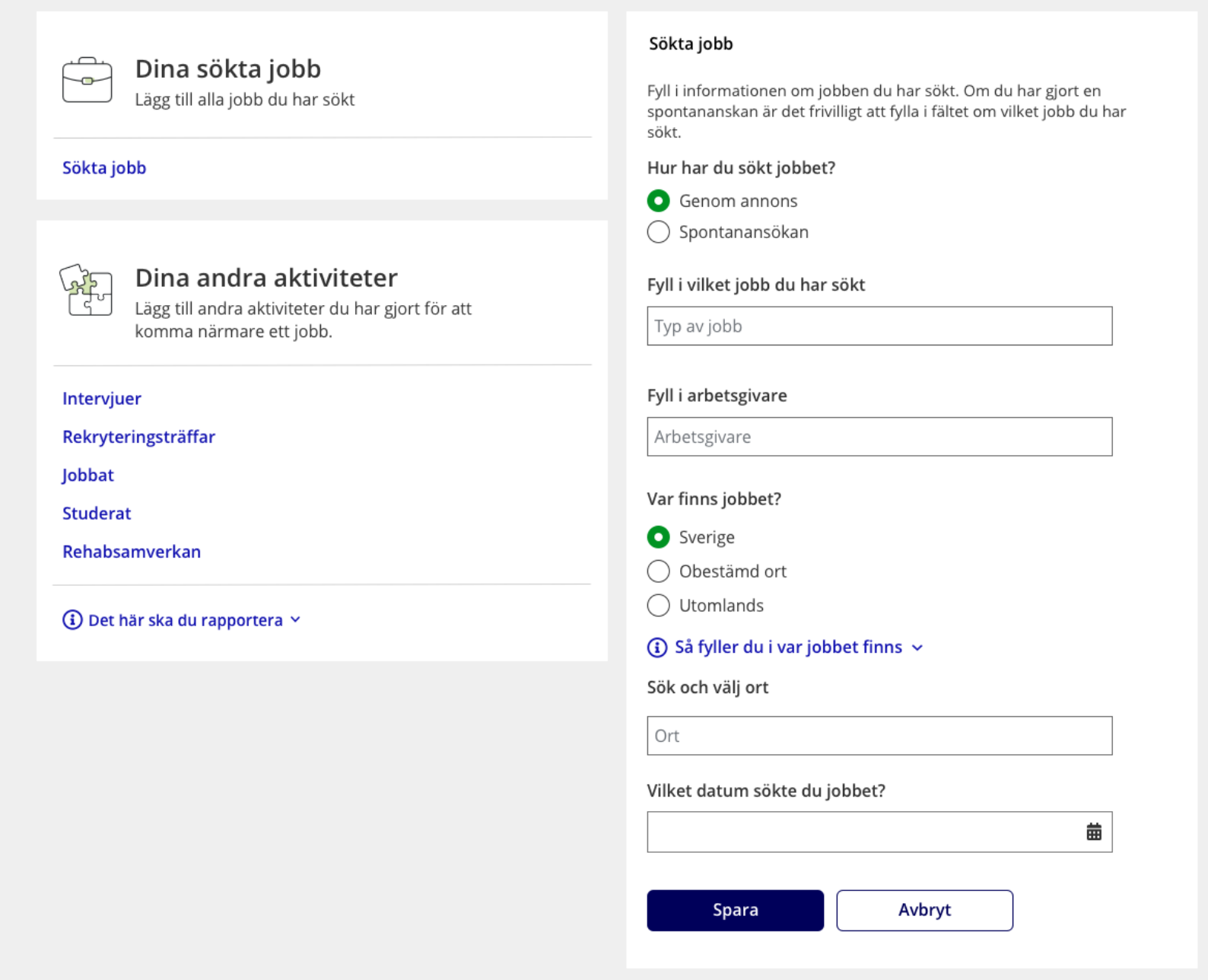} 
    \end{subfigure}
 \footnotesize Notes: The figure shows an example of the online interface (in Swedish) of the activity report from the Swedish PES website. 
\end{figure}

\begin{figure}[htbp]
  \caption{Distribution of applied-for jobs and interviews per month}
    \label{fig:hist}
     \begin{subfigure}[a]{.49\textwidth}
     \centering 
    \caption{\# of applied-for jobs}
     \label{fig:hista}
      \includegraphics[width=\textwidth]{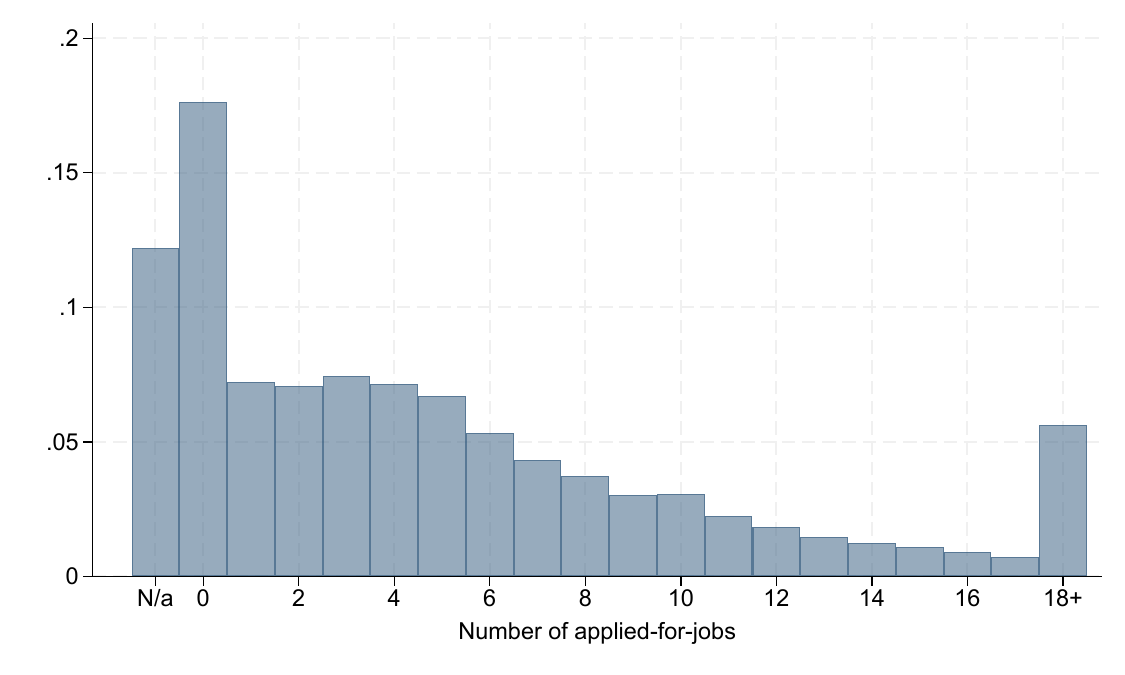}
    \end{subfigure}
    \begin{subfigure}[a]{.49\textwidth}
     \centering
    \caption{\# of interviews}
    \label{fig:histb}
      \includegraphics[width=\textwidth]{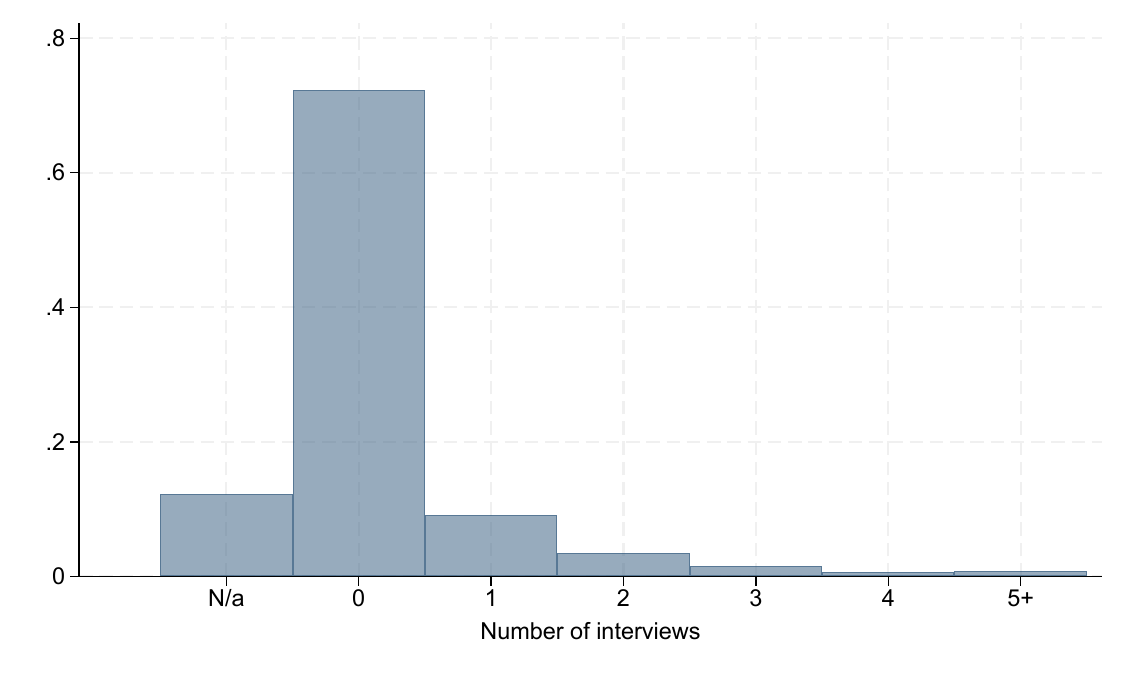}
    \end{subfigure}
    \footnotesize Notes: The figure shows for a given month the distribution of the number of applied-for jobs (panel a) and number of attended interviews (panel b), where we have exuded the first and last two calendar months of an unemployment spell. N/a refers to months where a job seeker has not submitted an activity report.
\end{figure}

\clearpage \newpage

\begin{figure}[htbp]
    \caption{Employment and earnings relative to having found a job} 
    \label{fig:earfoundjob}
     \begin{subfigure}[a]{.49\textwidth}
     \centering 
    \caption{Earnings}
     \label{fig:earfoundjoba}
      \includegraphics[width=\textwidth]{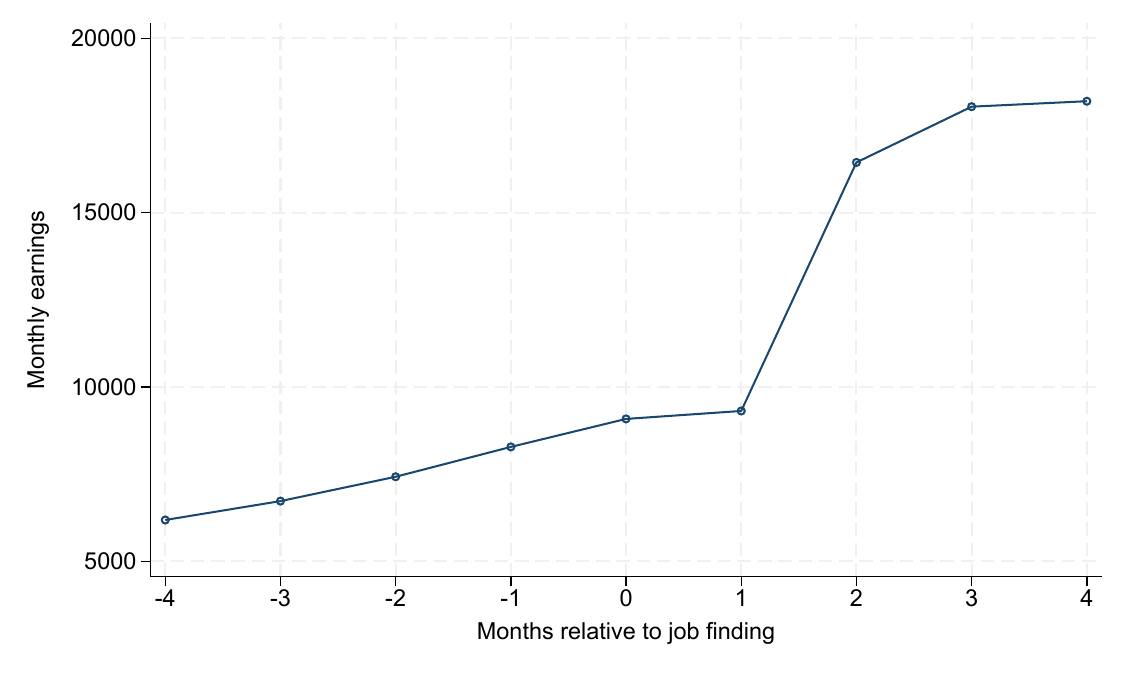}
    \end{subfigure}
    \begin{subfigure}[a]{.49\textwidth}
     \centering
    \caption{Pr(employment)}
    \label{fig:earfoundjobb}
      \includegraphics[width=\textwidth]{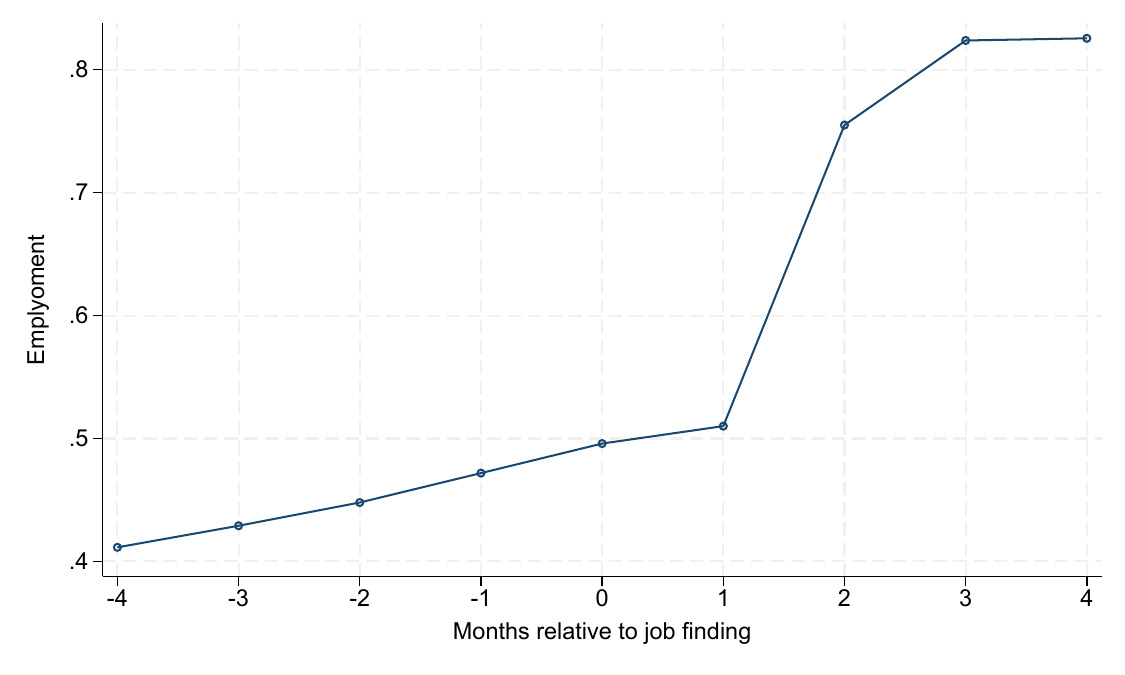}
    \end{subfigure}
    \footnotesize Notes: The figure shows how labor earnings (panel a) and probability of employment (panel b) by months relative to two months prior to end of the unemployment spell which we define as the time of having found a job. 
\end{figure}

\begin{figure}[htbp]
    \caption{Employment and earnings relative to last interview in spell} 
    \label{fig:earlastinter}
     \begin{subfigure}[a]{.49\textwidth}
     \centering 
    \caption{Earnings}
     \label{fig:earlastintera}
      \includegraphics[width=\textwidth]{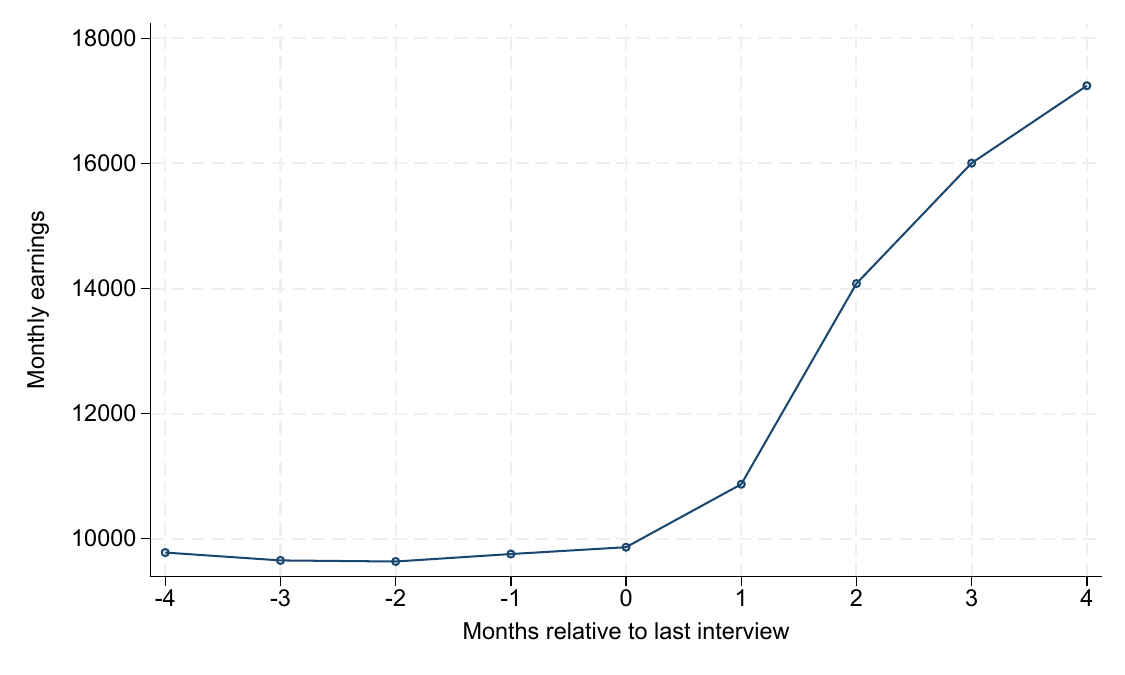}
    \end{subfigure}
    \begin{subfigure}[a]{.49\textwidth}
     \centering
    \caption{Pr(employment)}
    \label{fig:earlastinterb}
      \includegraphics[width=\textwidth]{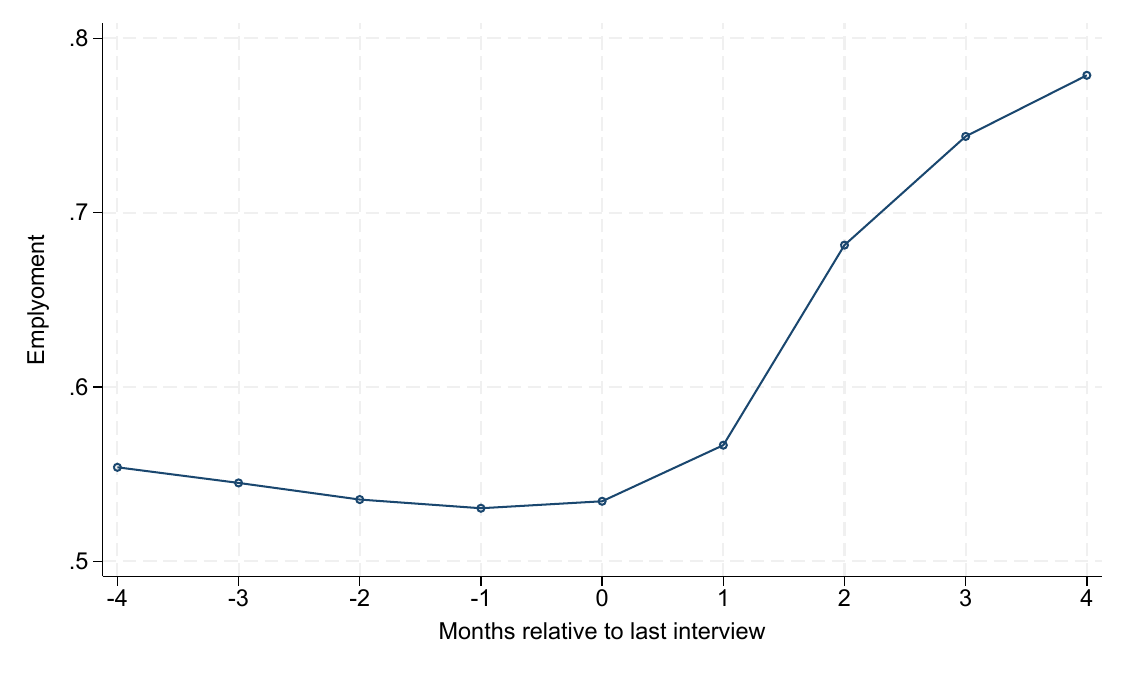}
    \end{subfigure}
    \footnotesize Notes: The figure shows how labor earnings (panel a) and probability of employment (panel b) by months relative to the last attended interview happening within an unemployment spell. 
\end{figure}

\clearpage \newpage

\section{Additional results on search effort} 
\label{app:appendixB}
\counterwithin{figure}{section}
\counterwithin{table}{section}
\counterwithin{equation}{section}
\setcounter{table}{0}
\setcounter{figure}{0}

\begin{figure}[htbp]
  \caption{Number of applied-for jobs by time until finding a job}
    \label{fig:searchback0}
     \begin{subfigure}[a]{.49\textwidth}
     \centering 
    \caption{Excluding last calendar month}
     \label{fig:searchback0a}
      \includegraphics[width=\textwidth]{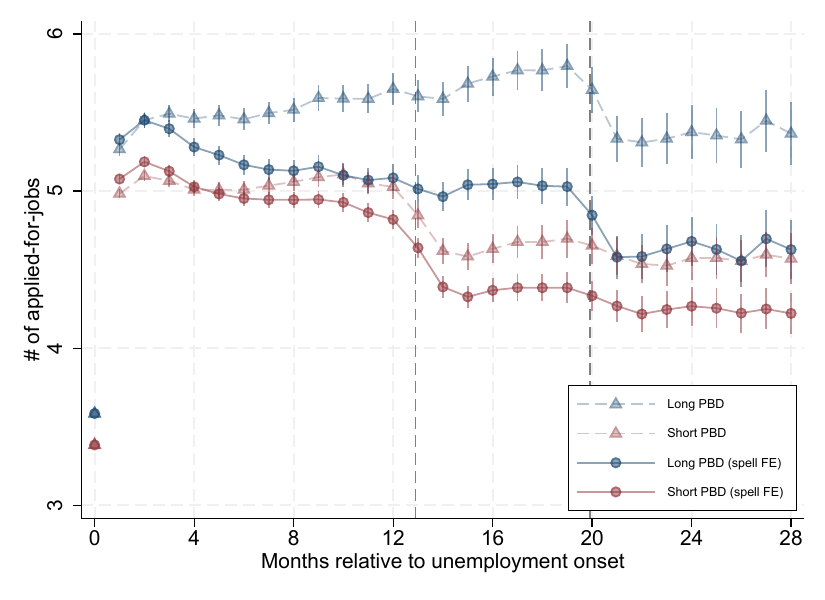}
    \end{subfigure}
    \begin{subfigure}[a]{.49\textwidth}
     \centering
    \caption{Excluding two last calendar months}
    \label{fig:searchback0b}
      \includegraphics[width=\textwidth]{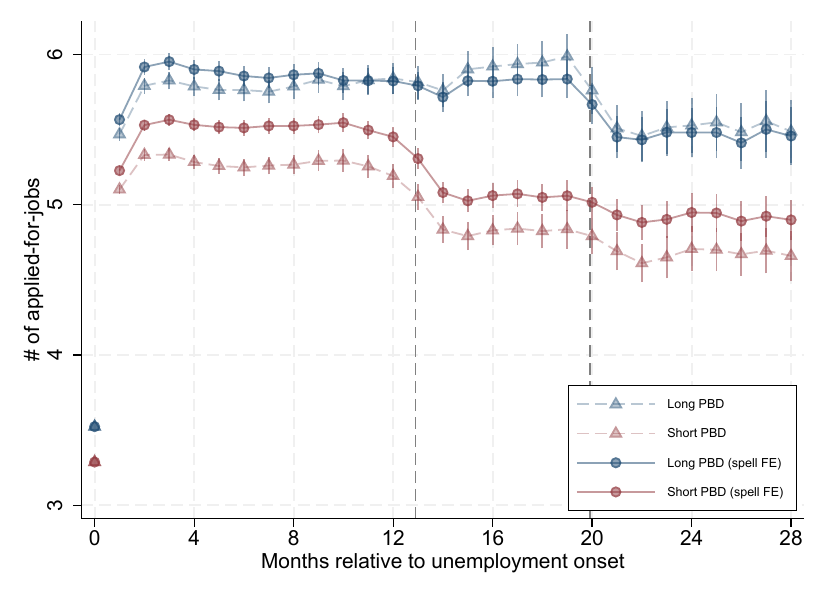}
    \end{subfigure} \\
    \footnotesize Notes: The figure shows the evolution of search intensity relative to unemployment onset for job seekers entitled to 14 (red) or 21 (blue) months of UI indicated by the dashed vertical lines. We censor the last two calendar months of each unemployment spell, considering search effort until finding a new job. The circles connected by dashed lines show estimates from equation \eqref{eq:timeFE} and reflect average number of applied-for jobs among the sample of job seekers remaining in unemployment in a given month. The triangles connected by solid lines depict estimates from \eqref{eq:spellFE}, accounting for dynamic selection and reflect within-spell changes in search intensity. Surrounding each estimate are 95\% confidence intervals with standard errors are clustered at the spell level.
\end{figure}

\begin{figure}[htbp]
  \caption{Hazard rate by unemployment duration} 
  \label{fig:hazard}
 \begin{subfigure}[a]{\textwidth}
 \centering 
 \includegraphics[width=.65\textwidth]{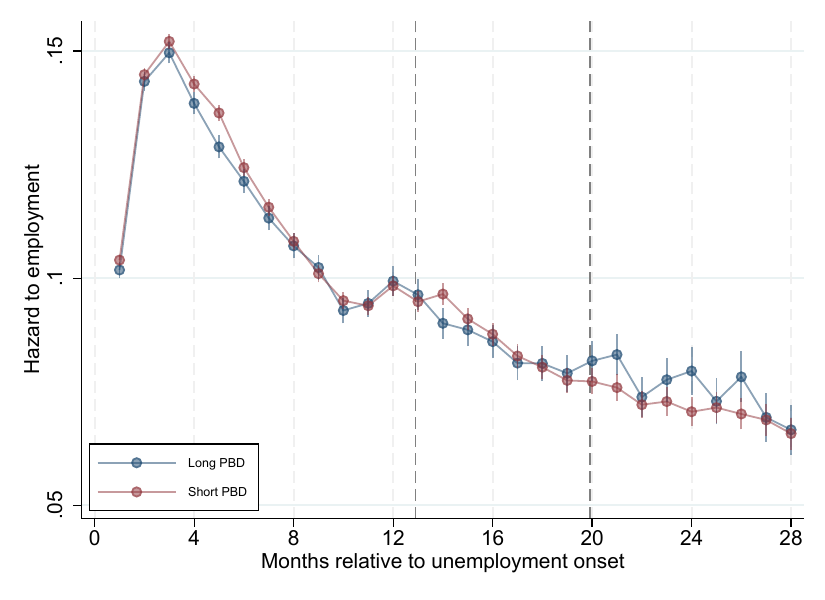}
\end{subfigure}
\footnotesize Notes: The figure shows estimates from equation \eqref{eq:timeFE} on the hazard to employment by unemployment duration estimated separately for job seekers eligible to 14 and 21 months of UI in red and blue, respectively. Surrounding each estimate are 95\% confidence intervals with standard errors are clustered at the spell level.
\end{figure}

\begin{figure}[t!]
  \caption{Evolution of other activities and search effort conditional on submitting report}
    \label{fig:otherint}
     \begin{subfigure}[a]{.49\textwidth}
     \centering 
    \caption{Pr(other activities)}
     \label{fig:otherinta}
      \includegraphics[width=\textwidth]{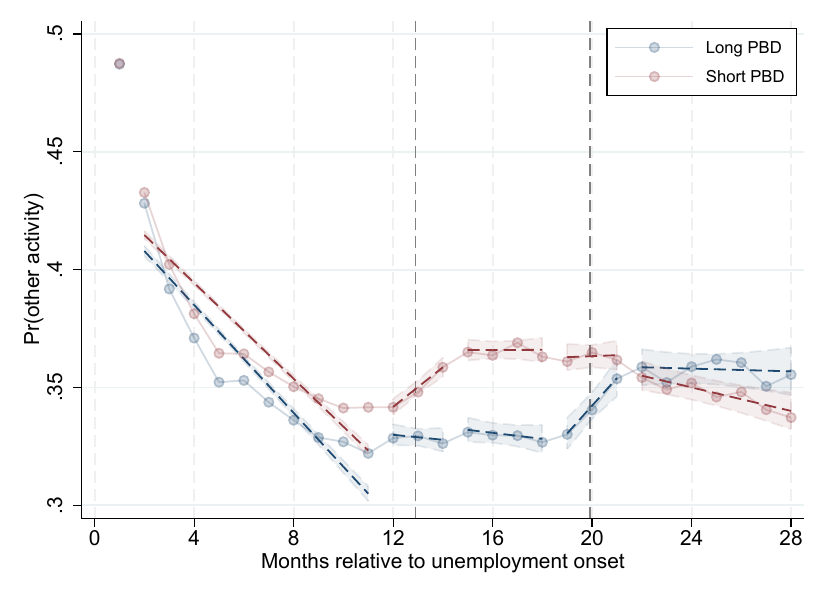}
    \end{subfigure}
    \begin{subfigure}[a]{.49\textwidth}
     \centering
    \caption{\# applied-for jobs }
    \label{fig:otherintb}
      \includegraphics[width=\textwidth]{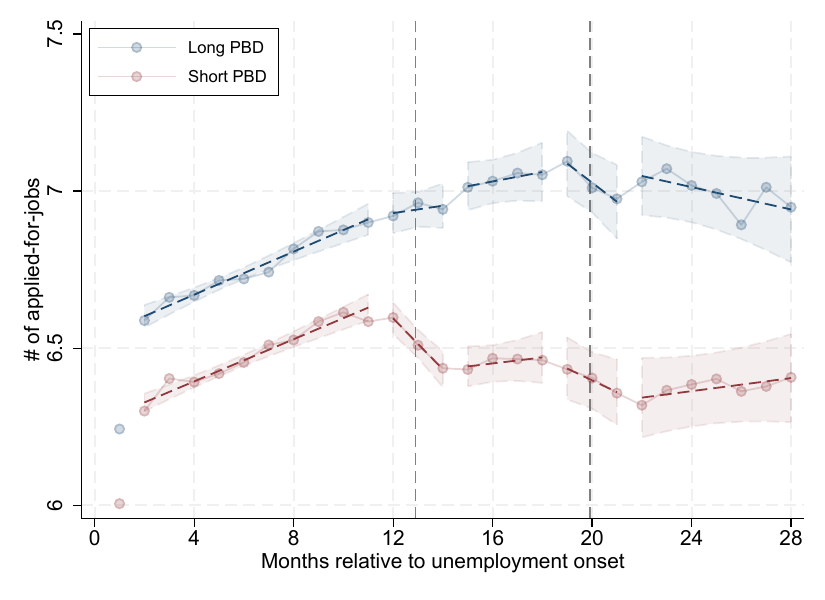}
    \end{subfigure} \\
    \footnotesize Notes: The figure shows within spell variation in the probability of reporting other activities aside from applied-for jobs (panel a) and the number of applied-for jobs conditional of having submitted an activity report. The blue and red dots show estimates from equation \ref{eq:spellFE} run separately for job seekers eligible to 21 and 14 months of PBD on UI, respectively, which are indicated by the vertical dashed line. The colored dashed lines show linear predictions along with 95\% confidence bands, having regressed number of applied-for jobs on elapsed duration, interacted with an indicator for different segments $k$ of duration $t$ being within the interval $k=\{[2,11],[12,14],[15,18],[19,21],[22,28]\}$. All spells are right-censored at the last two calendar months of each unemployment spell. Standard errors are clustered at the spell level.
\end{figure}

\begin{table}[htbp]\centering 
\begin{threeparttable}[b]
\def\sym#1{\ifmmode^{#1}\else\(^{#1}\)\fi} \small
\caption{Search effort by elapsed duration} \label{tab:otherint}
\begin{tabular}{l*{8}{c}}
\toprule
&\multicolumn{2}{c}{Short PBD}&\multicolumn{2}{c}{Long PBD} &\multicolumn{2}{c}{Short PBD}&\multicolumn{2}{c}{Long PBD} \\
\cmidrule(lr){2-3} \cmidrule(lr){4-5} \cmidrule(lr){6-7} \cmidrule(lr){8-9}  
&\multicolumn{4}{c}{Pr(other activities)}&\multicolumn{4}{c}{applied-for jobs per month}\\
& $ \Delta $Pr & $ \Delta $\%  & $ \Delta $Pr & $ \Delta $\%  & $ \Delta $\# & $ \Delta $\%  & $ \Delta $\# & $ \Delta $\%  \\
&(1) & (2) & (3) & (4) &(5) & (6) & (7) & (8) \\
\midrule 
\multicolumn{2}{l}{Duration (months)}&&&&&&\\
 
\quad 2--11 &      -0.010\sym{***}&      -2.350         &      -0.011\sym{***}&      -2.670         &       0.034\sym{***}&       0.530         &       0.034\sym{***}&       0.520         \\
            &     (0.000)         &                     &     (0.000)         &                     &     (0.003)         &                     &     (0.004)         &                     \\
 
\quad 12--14&       0.008\sym{***}&       2.460         &      -0.001         &      -0.320         &      -0.081\sym{***}&      -1.230         &       0.012         &       0.170         \\
            &     (0.001)         &                     &     (0.002)         &                     &     (0.015)         &                     &     (0.020)         &                     \\
 
\quad 15--18&       0.000         &       0.010         &      -0.001         &      -0.380         &       0.009         &       0.150         &       0.015         &       0.210         \\
            &     (0.001)         &                     &     (0.001)         &                     &     (0.014)         &                     &     (0.016)         &                     \\
 
\quad 19--21&       0.000         &       0.120         &       0.012\sym{***}&       3.560         &      -0.038         &      -0.580         &      -0.061\sym{**} &      -0.860         \\
            &     (0.002)         &                     &     (0.002)         &                     &     (0.025)         &                     &     (0.030)         &                     \\
 
\quad 22--28&      -0.002\sym{***}&      -0.700         &      -0.000         &      -0.080         &       0.010         &       0.160         &      -0.018         &      -0.250         \\
            &     (0.001)         &                     &     (0.001)         &                     &     (0.012)         &                     &     (0.015)         &                     \\
\cmidrule(lr){2-3} \cmidrule(lr){4-5} \cmidrule(lr){6-7} \cmidrule(lr){8-9}
Mean outcome ($ t$=2) &\multicolumn{2}{c}{0.43}&\multicolumn{2}{c}{0.43} &\multicolumn{2}{c}{6.30}&\multicolumn{2}{c}{6.59}\\
\# observations&\multicolumn{2}{c}{1,501,295}&\multicolumn{2}{c}{849,605}&\multicolumn{2}{c}{1,288,720}&\multicolumn{2}{c}{731,800}\\
\bottomrule
\end{tabular}
\begin{tablenotes}[flushleft]
\scriptsize \item \noindent  Notes: The table shows the average per month within-spell change in the probability of reporting other job search activities aside from applying for jobs (columns 1--4) and the number of applied-for jobs conditional on having submitted an activity report, by different segments of elapsed duration. The estimates come from the regression: $ y_{it} = \alpha_i + \sum_{k} \left[ \gamma_k (D_k \times t) + \delta_k D_k \right] + \varepsilon_{it}$ where $t$ denotes months in unemployment, and $D_k$ is an indicator  equal to one if $t$ falls within interval $k \in \{[2,11], [12,14], [15,18], [19,21], [22,28]\}$ and 0 otherwise. Columns (2), (4), (6) and (8) show the implied average percentage change per month calculated by plugging in the predictions from the ends of each interval $k\in[\underline{k},\overline{k}]$ into the compound decay formula: $ \Delta_k\% = 1 - \left(\hat{y}_{\overline{k}}/\hat{y}_{\underline{k}} \right)^{\frac{1}{\overline{k} - \underline{k}}}$. All spells are right-censored at the last two calendar months of each unemployment spell. Standard errors are in parenthesis and clustered at the spell level. Asterisks indicate that the estimates are significantly different from zero at the  \sym{*} \(p<0.1\), \sym{**} \(p<0.05\), \sym{***}  \(p<0.01\) level.
\end{tablenotes}
\end{threeparttable}
\end{table}

\begin{figure}[t!]
  \caption{Heterogeneity in search effort by unemployment duration}
    \label{fig:hetrosearch}
     \begin{subfigure}[a]{.49\textwidth}
     \centering 
    \caption{Native vs. non-native}
     \label{fig:hetrosearcha}
      \includegraphics[width=\textwidth]{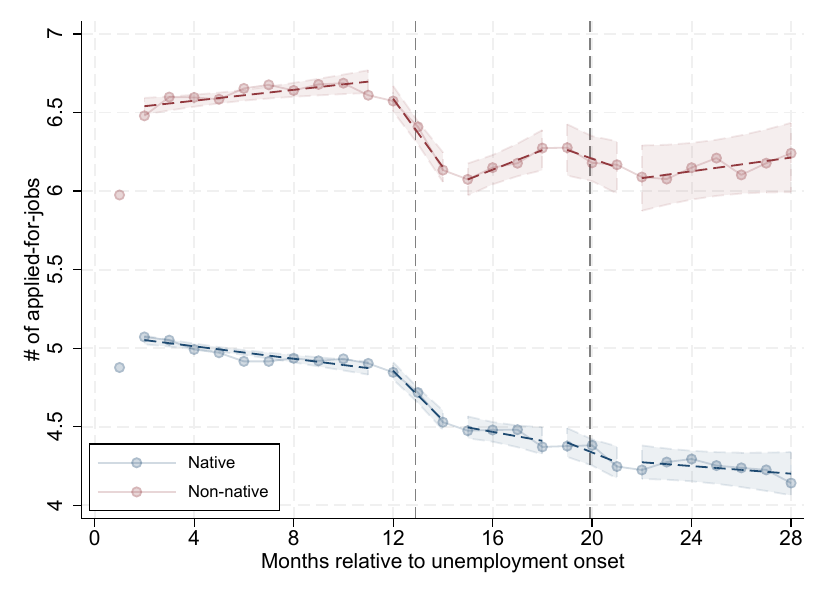}
    \end{subfigure}
    \begin{subfigure}[a]{.49\textwidth}
     \centering
    \caption{Male vs. Female}
    \label{fig:hetrosearchb}
      \includegraphics[width=\textwidth]{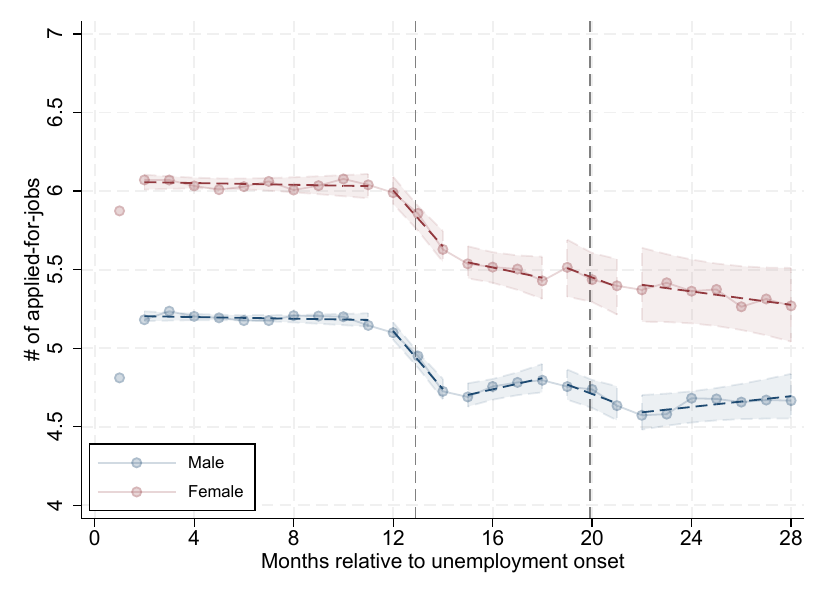}
    \end{subfigure} \\
     \begin{subfigure}[a]{.49\textwidth}
     \centering 
    \caption{High-skilled vs. low-skilled}
     \label{fig:hetrosearchc}
      \includegraphics[width=\textwidth]{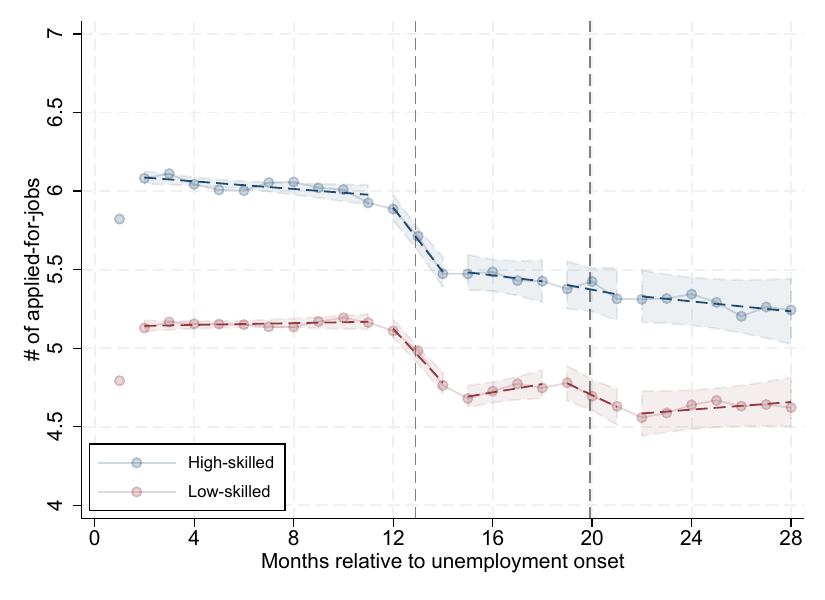}
    \end{subfigure}
    \begin{subfigure}[a]{.49\textwidth}
     \centering
    \caption{Tight vs. slack labor market}
    \label{fig:hetrosearchd}
      \includegraphics[width=\textwidth]{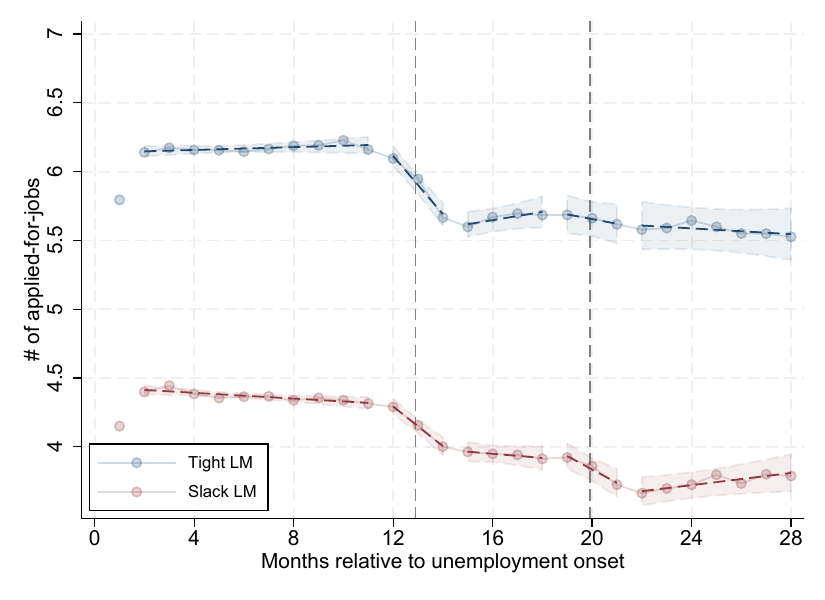}
    \end{subfigure}
    \footnotesize Notes: The Figure shows the evolution of within-spell search intensity relative to unemployment onset for job seekers entitled to 14 months of UI. Estimates come from estimating equation \eqref{eq:spellFE} separately for each group of job seekers. The circles represent average within-spell search effort in each month whereas the dashed colored lines show linear predictions in different segments of elapsed duration. Surrounding the linear predictions are 95\% confidence bands with standard errors are clustered at the spell level. Panel a) show search intensity for native (blue) and non-native (red) job seekers whereas panel b) show male (blue) and female (red) job seekers. Panel c) show estimates for job seeker targeting high (blue) and low-skilled (red) occupations as defined by their stated occupational preferences at the beginning of the unemployment spell. Panel d) displays estimates for spell ongoing in tight (blue) and slack (red) local labor markets, defined by the average regional$\times$quarterly vacancy rate being above and below the median in the distribution, respectively. In all regressions we right-censor the last two calendar months of each unemployment spell, considering duration dependence until finding a new job.
\end{figure}

  \clearpage \newpage

\section{Additional results on duration dependence} 
\label{app:appendixC}
\counterwithin{figure}{section}
\counterwithin{table}{section}
\counterwithin{equation}{section}
\setcounter{table}{0}
\setcounter{figure}{0}

\begin{figure}[htbp]
  \caption{Pr(interview) by unemployment duration separately by PBD} 
  \label{fig:interviewPBD}
 \begin{subfigure}[a]{\textwidth}
 \centering 
 \includegraphics[width=.65\textwidth]{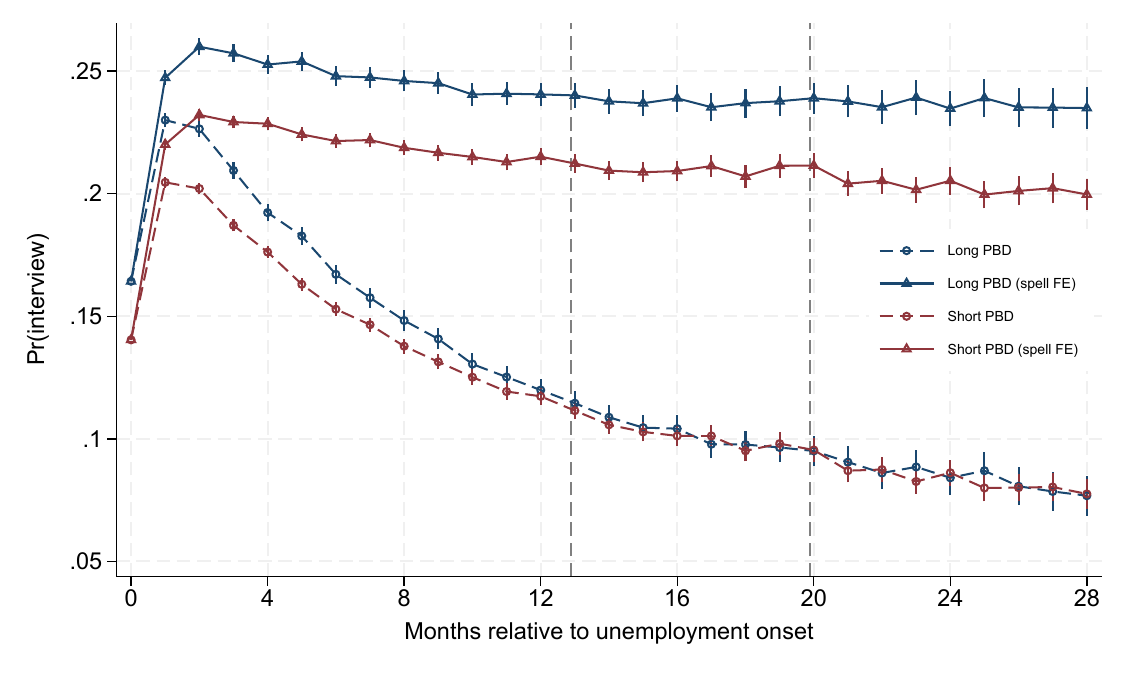}
\end{subfigure}
\footnotesize Notes: Notes: The figure shows the evolution of the probability of attending an interview relative to unemployment onset. The circles connected by dashed lines show estimates from equation \eqref{eq:timeFE} and reflect average probability of attending an interview in a given month among the sample of job seekers remaining in unemployment. The triangles connected by solid lines depict estimates from \eqref{eq:spellFE}, accounting for dynamic selection and reflect within-spell changes, i.e true duration dependence in callbacks. Spells are right-censored in the last two calendar months of each unemployment spell, considering search effort until finding a new job. Surrounding each estimate are 95\% confidence intervals with standard errors are clustered at the spell level.
\end{figure}

\begin{figure}[t]
  \caption{Pr(interview) by fixed unemployment durations} 
  \label{fig:interviewfixed}
 \begin{subfigure}[a]{\textwidth}
 \centering 
 \includegraphics[width=.65\textwidth]{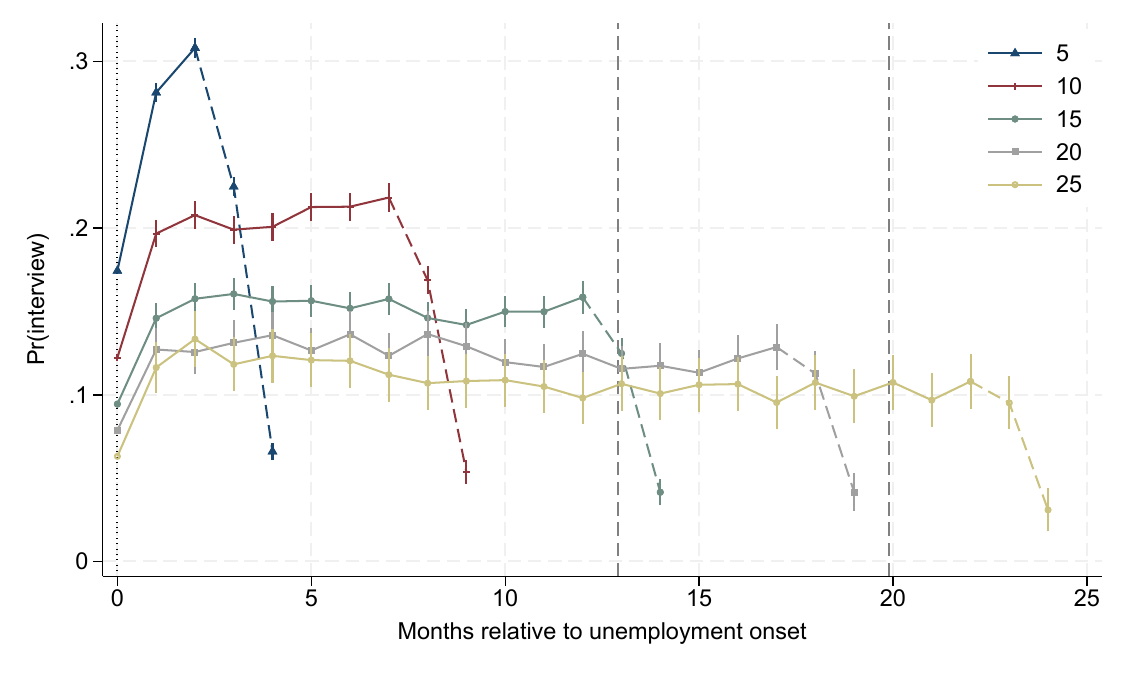}
\end{subfigure}
\footnotesize Notes: The figure shows the evolution of attending an interview relative to unemployment onset for job seekers with 5, 10, 15, 20 or 25 months of realized unemployment duration, estimated using the model in equation \eqref{eq:timeFE}. These estimates pool job seekers with different PBD. The dashed portion of each solid line reflect the last two calendar months in the unemployment spell which we view as being the interim period between securing employment and starting a new job. Surrounding each estimate are 95\% confidence intervals with standard errors are clustered at the spell level.
\end{figure}

\begin{figure}[t]
  \caption{Poisson estimates of percentage change in Pr(interview) } 
  \label{fig:poisson}
 \begin{subfigure}[a]{\textwidth}
 \centering 
 \includegraphics[width=.75\textwidth]{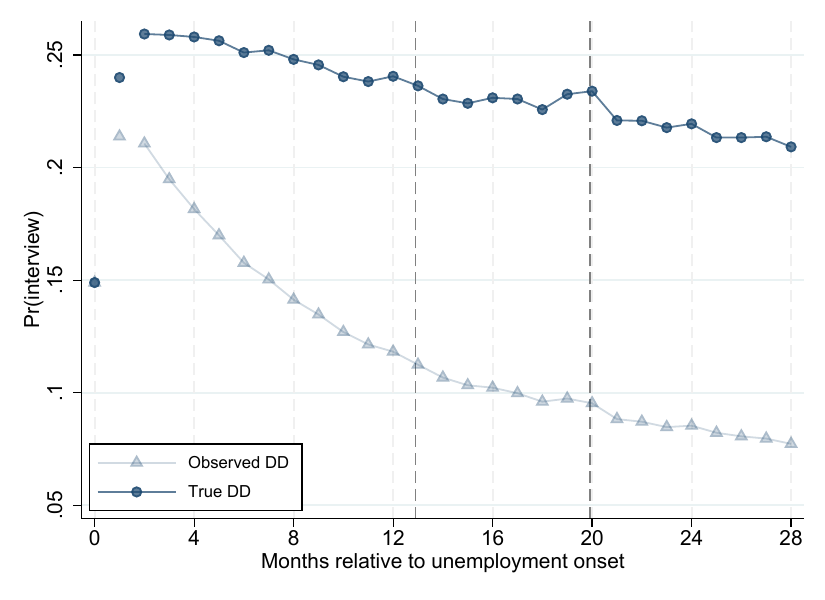}
\end{subfigure}
\footnotesize Notes: The figure shows the evolution of the probability of attending an interview relative to onset of unemployment. The light colored triangles show estimates from a Poisson regression of equation \eqref{eq:timeFE} reflecting \textit{observed} duration dependence whereas the circles connected by solid lines depict estimates from a Poisson regression og equation \eqref{eq:spellFE} which accounts for dynamic selection, thus reflecting \textit{true} duration dependence. We censor the last two calendar months of each unemployment spell, considering duration dependence until finding a new job. 
\end{figure}

\begin{table}[htbp]\centering \small
\begin{threeparttable}[b]
\def\sym#1{\ifmmode^{#1}\else\(^{#1}\)\fi}
\caption{Estimates of duration dependence controlling for past search effort} \label{tab:DDmainsearch}
\begin{tabular}{l*{4}{c}}
\toprule
&\multicolumn{2}{c}{Linear model}&\multicolumn{2}{c}{Saturated model} \\
&w/o spell-FE & w. spell-FE &w/o spell-FE & w. spell-FE \\
\cmidrule(lr){2-3} \cmidrule(lr){4-5}  
&(1) & (2) & (3) & (4)  \\
\midrule  
\multicolumn{1}{l}{\textbf{Panel a): Within 12 months}} &&& \\
Elapsed duration&     -0.0101\sym{***}&         -0.0013\sym{***}&             &            \\
            &    (0.0001)&         (0.0001)&            &            \\
            &  [-6.641\%]&  [-0.743\%]&  [-5.892\%]&  [-0.560\%]\\
\addlinespace            
  $ \widehat{\tau}_{2}: \widehat{Pr}(interview)_{t+2} $ &       0.203&       0.145&       0.213&       0.169\\
            &     (0.002)&     (0.002)&     (0.003)&     (0.003)\\
  $ \widehat{\tau}_{12}: \widehat{Pr}(interview)_{t+12} $ &       0.102&       0.135&       0.116&       0.160\\
            &     (0.002)&     (0.002)&     (0.003)&     (0.003)\\
\cmidrule(lr){2-3} \cmidrule(lr){4-5}
Share of true dependence&   \multicolumn{2}{c}{11.2\%}&   \multicolumn{2}{c}{9.5\%}\\
\# observations&   \multicolumn{2}{c}{1,339,238}&   \multicolumn{2}{c}{1,339,238}\\
\addlinespace
\multicolumn{1}{l}{\textbf{Panel b): Within 24 months}} &&& \\
Elapsed duration&     -0.0064\sym{***}&         -0.0009\sym{***}&            &            \\
            &    (0.0001)&         (0.0001)&               &            \\
            &  [-5.947\%]&  [-0.622\%]&   [4.027\%]&  [-0.492\%]\\
\addlinespace            
  $ \widehat{\tau}_{2}: \widehat{Pr}(interview)_{t+2} $ &       0.191&       0.146&       0.213&       0.169\\
            &     (0.002)&     (0.002)&     (0.003)&     (0.003)\\
  $ \widehat{\tau}_{24}: \widehat{Pr}(interview)_{t+24} $ &       0.050&       0.128&       0.086&       0.152\\
            &     (0.002)&     (0.002)&     (0.003)&     (0.003)\\
\cmidrule(lr){2-3} \cmidrule(lr){4-5}
Share of true dependence&   \multicolumn{2}{c}{10.5\%}&   \multicolumn{2}{c}{12.2\%}\\
\# observations&   \multicolumn{2}{c}{1,722,770}&   \multicolumn{2}{c}{1,722,770}\\
\bottomrule
\end{tabular}
\begin{tablenotes}[flushleft]
\scriptsize \item \noindent  Notes: The table shows how the probability of attending an interview evolves during within 12 (panel a) and 24 (panel b) months of unemployment while holding constant lagged search effort. Columns (1) and (2) show estimates from regressing a dummy of having a attended an interview on a continuous variable of elapsed unemployment duration, with and without spell-fixed effects, respectively. Columns (3) and (4) show $\tau$-estimates from equation \eqref{eq:timeFE} and \eqref{eq:spellFE}, respectively. In hard brackets are the implied average percentage change per month calculated with compound decay formula: $ \Delta_k\% = 1 - \left(\hat{y}_{\overline{k}}/\hat{y}_{\underline{k}} \right)^{\frac{1}{\overline{k} - \underline{k}}}$ where $k \in \{[2,12], [2,24]\}$. All spells are right-censored at the last two calendar months of each unemployment spell. Standard errors are in parenthesis and clustered at the spell level. Asterisks indicate that the estimates are significantly different from zero at the  \sym{*} \(p<0.1\), \sym{**} \(p<0.05\), \sym{***}  \(p<0.01\) level.
\end{tablenotes}
\end{threeparttable}
\end{table}

\begin{figure}[t!]
  \caption{Heterogeneity in duration dependence}
    \label{fig:hetrodur}
     \begin{subfigure}[a]{.49\textwidth}
     \centering 
    \caption{Native vs. non-native}
     \label{fig:hetrodura}
      \includegraphics[width=\textwidth]{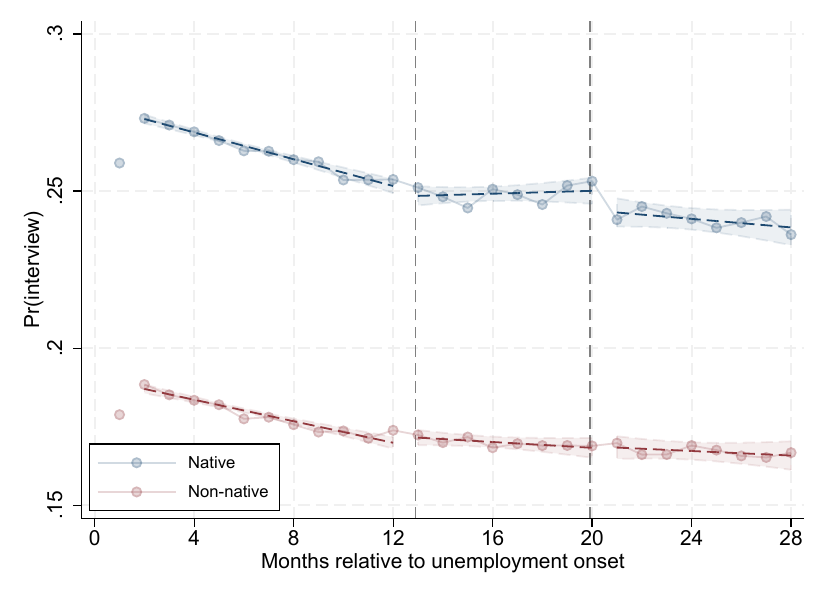}
    \end{subfigure}
    \begin{subfigure}[a]{.49\textwidth}
     \centering
    \caption{Male vs. Female}
    \label{fig:hetrodurb}
      \includegraphics[width=\textwidth]{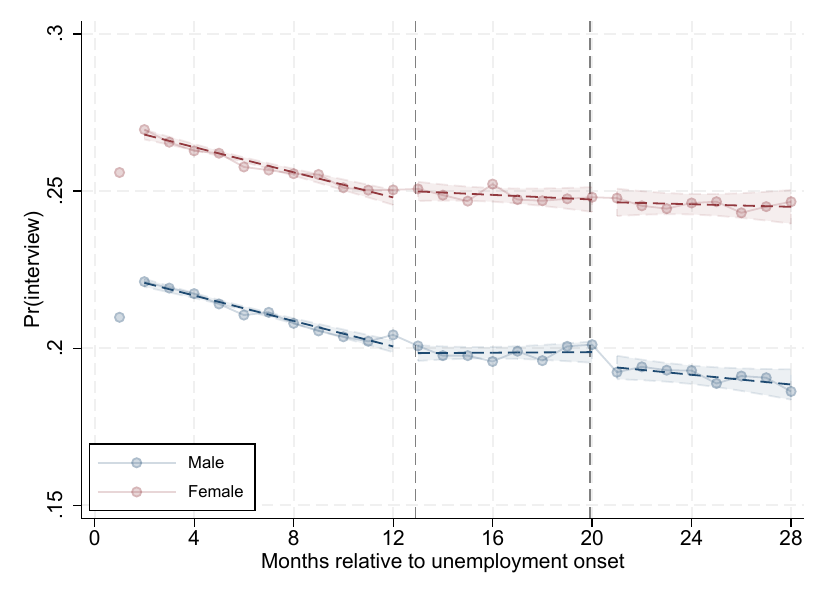}
    \end{subfigure} \\
     \begin{subfigure}[a]{.49\textwidth}
     \centering 
    \caption{High-skilled vs. low-skilled}
     \label{fig:hetrodurc}
      \includegraphics[width=\textwidth]{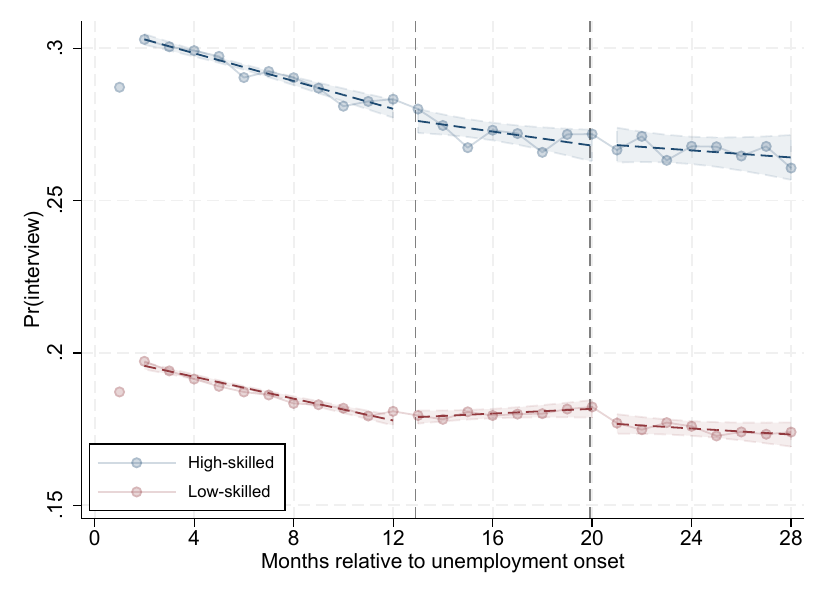}
    \end{subfigure}
    \begin{subfigure}[a]{.49\textwidth}
     \centering
    \caption{Tight vs. slack labor market}
    \label{fig:hetrodurd}
      \includegraphics[width=\textwidth]{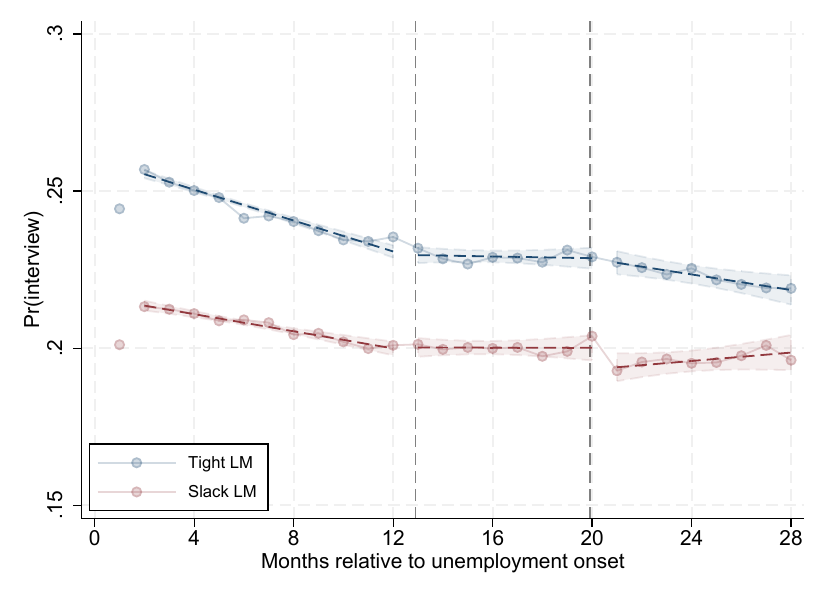}
    \end{subfigure}
    \footnotesize Notes: The Figure shows the evolution of getting called for an interview relative to unemployment onset for job seekers entitled to 14 months of UI. Estimates come from estimating equation \eqref{eq:spellFE} separately for each group of job seekers. The circles represent average within-spell estimates in each month whereas the dashed colored lines show linear predictions in different segments of elapsed duration. Surrounding the linear predictions are 95\% confidence bands with standard errors are clustered at the spell level. Panel a) show search intensity for native (blue) and non-native (red) job seekers whereas panel b) show male (blue) and female (red) job seekers. Panel c) show estimates for job seeker targeting high (blue) and low-skilled (red) occupations as defined by their stated occupational preferences at the beginning of the unemployment spell. Panel d) displays estimates for spell ongoing in tight (blue) and slack (red) local labor markets, defined by the average regional$\times$quarterly vacancy rate being above and below the median in the distribution, respectively. In all regressions we censor the last two calendar months of each unemployment spell, considering duration dependence until finding a new job.
\end{figure}

\begin{figure}[t!]
  \caption{Heterogeneity in duration dependence}
    \label{fig:hetrodurmean}
     \begin{subfigure}[a]{.49\textwidth}
     \centering 
    \caption{Native vs. non-native}
     \label{fig:hetrodurmeana}
      \includegraphics[width=\textwidth]{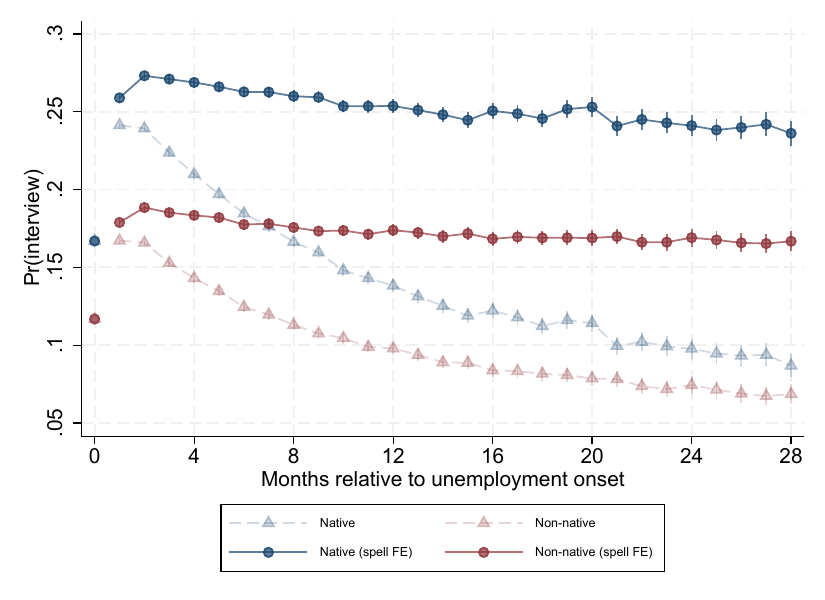}
    \end{subfigure}
    \begin{subfigure}[a]{.49\textwidth}
     \centering
    \caption{Male vs. Female}
    \label{fig:hetrodurmeanb}
      \includegraphics[width=\textwidth]{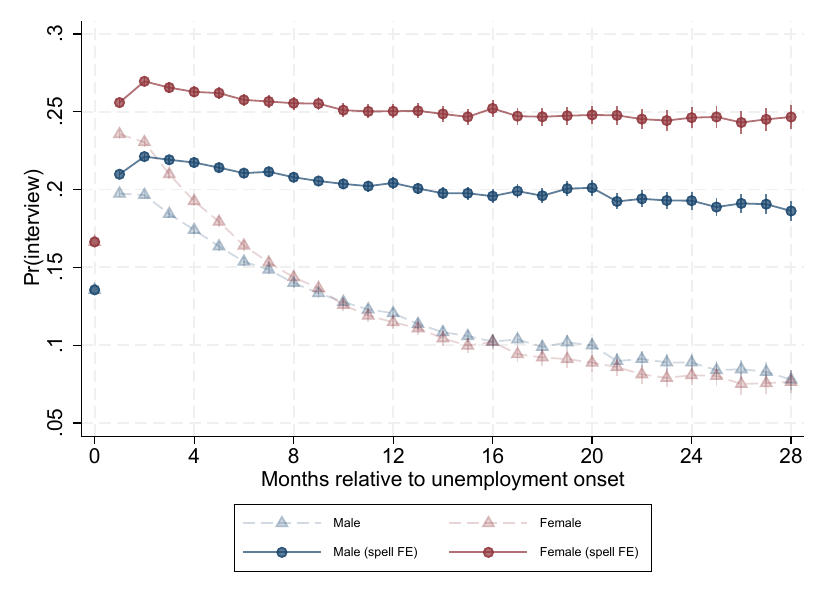}
    \end{subfigure} \\
     \begin{subfigure}[a]{.49\textwidth}
     \centering 
    \caption{Low-skilled vs. high-skilled}
     \label{fig:hetrodurmeanc}
      \includegraphics[width=\textwidth]{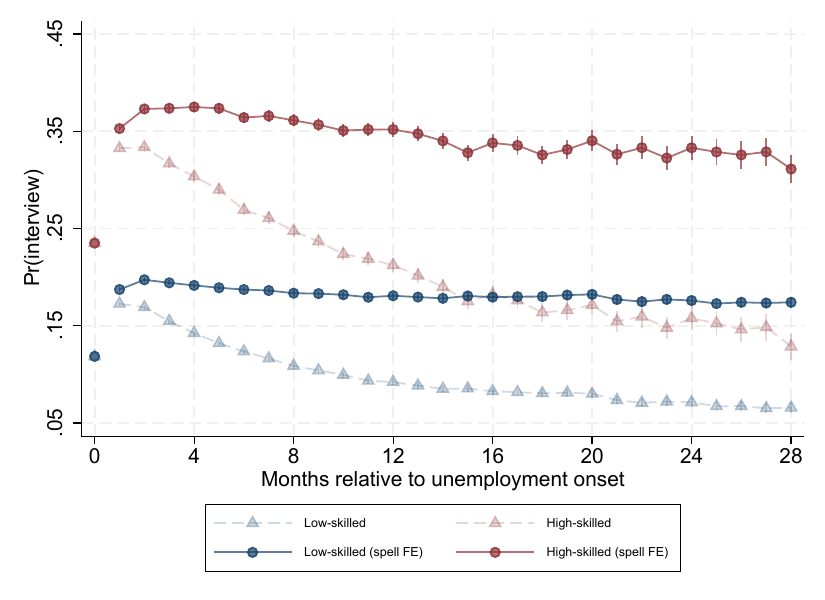}
    \end{subfigure}
    \begin{subfigure}[a]{.49\textwidth}
     \centering
    \caption{Tight vs. slack labor market}
    \label{fig:hetrodurmeand}
      \includegraphics[width=\textwidth]{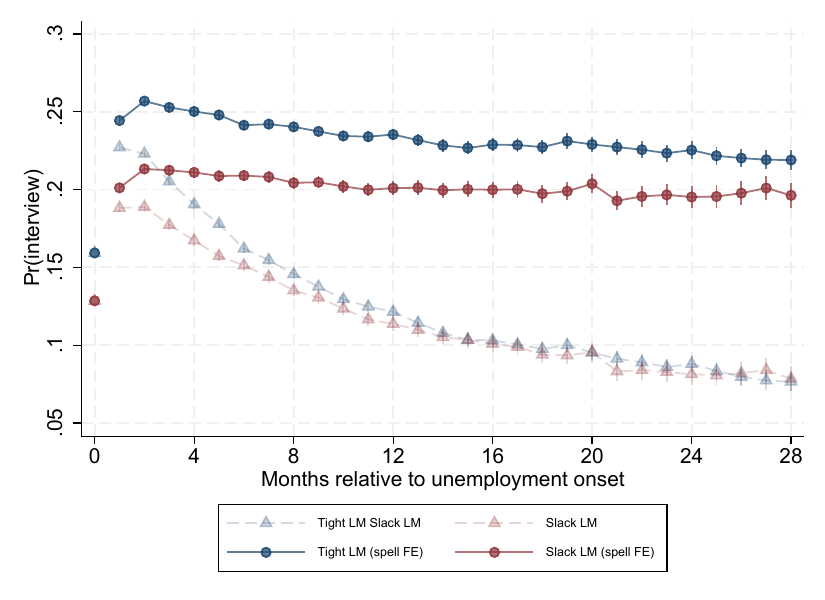}
    \end{subfigure}
    \footnotesize Notes: The figure shows the evolution of the probability of attending an interview relative to onset of unemployment. The circles connected by dashed lines show estimates from equation \eqref{eq:timeFE} reflecting \textit{observed} duration dependence whereas the triangles connected by solid lines depict estimates from \eqref{eq:spellFE} which accounts for dynamic selection, thus reflecting \textit{true} duration dependence. Panel a) shows separate estimates for native (blue) and non-native (red) job seekers and in panel b) separate estimates for male (blue) and female (red) job seekers. Panel c) depicts estimates for job seeker targeting high (red) and low-skilled (blue) occupations as defined by their stated occupational preferences at the beginning of the unemployment spell. Panel d) displays estimates for spell ongoing in slack (red) and tight (blue) local labor markets, defined by the average regional$\times$quarterly vacancy rate being below and above the median in the distribution, respectively. In all regressions we censor the last two calendar months of each unemployment spell, considering duration dependence until finding a new job. Surrounding each estimate are 95\% confidence intervals with standard errors are clustered at the spell level.
\end{figure}

\begin{figure}[t]
  \caption{Duration dependence by age with and w/o controlling for search effort} 
  \label{fig:DDhetroage_cont_search}
 \begin{subfigure}[a]{\textwidth}
 \centering 
 \includegraphics[width=.65\textwidth]{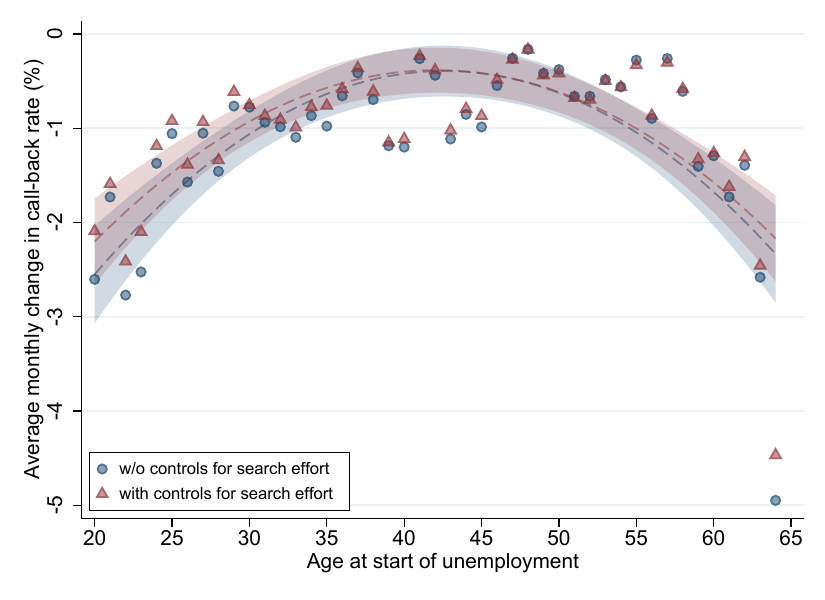}
\end{subfigure}
\footnotesize Notes: The figure shows the within-spell average month-to-month percent change, in the probability of attending an interview during the first year of unemployment separately estimated for each age cohort. The red triangles controls for an age-specific linear trend and levels of search effort whereas the blue dots do not. Estimates come from regressing a dummy of having a attended an interview on a continuous variable of elapsed unemployment duration, including time and spell fixed effects. We then take the linear predictions of this model and calculate the implied average percentage change per month using the compound decay formula: $\Delta_k\% = 1 - \left(\hat{y}_{\overline{k}}/\hat{y}_{\underline{k}} \right)^{\frac{1}{\overline{k} - \underline{k}}}$ where $k \in\{2,12\}$. The dashed lines show a uniformly weighted quadratic fit over the estimates along with a 95\% prediction interval. All spells are right-censored at the last two calendar months of each unemployment spell.
\end{figure}

\clearpage

\section{Derivation of the mapping between the share of duration dependence in callbacks and job-finding} 
\label{app:appendixD}
\counterwithin{figure}{section}
\counterwithin{table}{section}
\counterwithin{equation}{section}
\setcounter{table}{0}
\setcounter{figure}{0}

We start by defining the job-finding rate ($h$) for a job seeker as follows:
\begin{equation}
\label{eqAPP:indhaz}
	h(x,t) = \lambda_o(x,t)\lambda_i(x,t)
\end{equation}
where $\lambda_o$ the job-offer arrival rate conditional on getting an interview and $\lambda_i(x,t)$ is the probability of being called to an interview. These are both a functions of elapsed time in unemployment $t$ and job seeker characteristics $x\in \mathcal{X}$ at time $t$ with a cross-sectional density function $f(x;t)$ normalized such that $\int_{\mathcal X} f(x;t)\,dx=1$. 

We define the cross-sectional averages in the hazard rate, job offer arrival rate and callback probability, respectively, as follows:
\[
\bar h(t)=\int_{\mathcal X} h(x,t)\,f(x;t)\,dx,\qquad
\bar\lambda_o(t)=\int_{\mathcal X}\lambda_o(x,t)\,f(x;t)\,dx \qquad
\bar\lambda_i(t)=\int_{\mathcal X}\lambda_i(x,t)\,f(x;t)\,dx.
\]
Let $\partial_t$ denote the derivative with respect to $t$. Differentiating the above equations we get:
\begin{align}
\frac{\partial h(x,t)}{\partial t}
&= (\partial_t\lambda_o(x,t))\,\lambda_i(x,t)
+ \lambda_o(x,t)\,(\partial_t\lambda_i(x,t)),
\\[4pt]
\frac{d\bar h(t)}{dt}
&= \int_{\mathcal X}\Big[(\partial_t\lambda_o)\lambda_i f
+ \lambda_o(\partial_t\lambda_i) f
+ \lambda_o\lambda_i(\partial_t f)\Big]\,dx,
\\[4pt]
\frac{d\bar\lambda_i(t)}{dt}
&= \int_{\mathcal X}\Big[(\partial_t\lambda_i) f
+ \lambda_i(\partial_t f)\Big]\,dx.
\end{align}
which are true duration dependence in the hazard rate, observed duration dependence in the hazard rate and the observed duration dependence in the interview probability, respectively. Dividing equation D.2 by D.3, we get the share of true duration dependence in in the job-finding rate:
\begin{equation}
	\label{eqAPP:shareDD}
	\mathcal R_h(x,t) \equiv
\frac{\partial h(x,t)/\partial t}{\,d\bar h(t)/dt\,}
=
\frac{(\partial_t\lambda_o)\lambda_i+\lambda_o(\partial_t\lambda_i)}
{\displaystyle \int_{\mathcal X}\!\Big[(\partial_t\lambda_o)\lambda_i f
+ \lambda_o(\partial_t\lambda_i) f
+ \lambda_o\lambda_i(\partial_t f)\Big]\,dx}
\end{equation}
The share of true duration dependence in callback rates is similarly defined by the ratio:
\begin{equation}
\label{eqAPP:shareCB}
\mathcal R_i(x,t) \equiv	\frac{\partial\lambda_i(x,t)/\partial t}{\,d\bar\lambda_i(t)/dt\,}
=
\frac{\partial_t \lambda_i(x,t)}
{\displaystyle \int_{\mathcal X}\!\Big[(\partial_t\lambda_i) f
+ \lambda_i(\partial_t f)\Big]\,dx}
\end{equation}

\noindent There are two conditions equation \eqref{eqAPP:shareDD} to be approximately equal to equation \eqref{eqAPP:shareCB} (i.e $\mathcal R_h(x,t) =\mathcal R_i(x,t) $).

Suppose the following conditions are met:
\begin{itemize}
\item[\textbf{(A1)}] There is weak heterogeneity in job-offer arrival rates such that $\lambda_o(x,t) \approx \bar{\lambda}_o(t)$ at any given duration $t$.
\item[\textbf{(A2)}] There is limited duration dependence in job-offer arrival rates such that $\partial_t \lambda_o \approx 0 $ \\
\end{itemize}
\vspace{-20pt}
Alternatively, one could replace condition (A2) with the somewhat less strict condition:
\begin{itemize}
\item[\textbf{(A2 alt.)}] There is limited duration dependence job-offer arrival rates relative that of callbacks both
	\begin{itemize}
	\item[(A2a)] at the individual level such that $(\partial_t\lambda_o)\lambda_i \ll \lambda_o(\partial_t\lambda_i)$, and \\
	\item[(A2b)] in the aggregate such that\\ $\int_{\mathcal X} (\partial_t \lambda_o(x,t))\,\lambda_i(x,t)\,f(x;t)\,dx \;\ll\; \int_{\mathcal X} \lambda_o(x,t)\,(\partial_t \lambda_i(x,t))\,f(x;t)\,dx$.
	\end{itemize}
\end{itemize}
\noindent Focusing on the denominator of \eqref{eqAPP:shareDD} we add and subtract $\bar{\lambda}_o(t)$ and note that it does not depend on $x$. The denominator can then be rewritten it as:
\begin{align}
&=	
\int_{\mathcal X}\!\Big[(\partial_t\lambda_o)\lambda_i f\Big]\,dx
+ \int_{\mathcal X}\!\lambda_o\Big[(\partial_t\lambda_i) f
+ \lambda_i(\partial_t f)\Big]\,dx \\[4pt]
&=\int_{\mathcal X}\!\Big[(\partial_t\lambda_o)\lambda_i f\Big]\,dx
+\int_{\mathcal X}\Big[\bar\lambda_o(t)+(\lambda_o-\bar\lambda_o(t))\Big]\Big[(\partial_t\lambda_i) f+\lambda_i(\partial_t f)\Big]\,dx \\
&=	
 \int_{\mathcal X}\!\Big[(\partial_t\lambda_o)\lambda_i f\Big]\,dx
+ \bar{\lambda}_o(t) \int_{\mathcal X}\! \Big[(\partial_t\lambda_i) f
+ \lambda_i(\partial_t f)\Big]\,dx  + \int_{\mathcal X}\! (\lambda_o - \bar{\lambda}_o(t))\Big[(\partial_t\lambda_i) f
+ \lambda_i(\partial_t f)\Big]\,dx.
\end{align}
\noindent Under assumption A1 the last term in is negligible and can be approximated as zero thus allowing the hazard ratio in equation \eqref{eqAPP:shareDD} to be re-written as:

\begin{equation}
	\frac{\partial h(x,t)/\partial t}{\,d\bar h(t)/dt\,}
\approx
\frac{(\partial_t\lambda_o)\lambda_i+\lambda_o(\partial_t\lambda_i)}
{\displaystyle \int_{\mathcal X}\!\Big[(\partial_t\lambda_o)\lambda_i f\Big]\,dx
+ \bar{\lambda}_o(t) \int_{\mathcal X}\! \Big[(\partial_t\lambda_i) f
+ \lambda_i(\partial_t f)\Big]\,dx} 
\end{equation}
Using assumption A2 that the job-offer arrival rate is either constant $\partial_t \lambda_o \approx 0 $, or under A2 alt. that it is small relative to the change in callbacks, the first terms in both the numerator and denominator can be ignored and the hazard ratio becomes:
\begin{equation}
\mathcal R_h(x,t) 
=	\frac{\partial_t h(x,t)/\partial t}{\,d\bar h(t)/dt\,}
\approx \frac{\lambda_o(\partial_t\lambda_i)}
{\displaystyle  \bar{\lambda}_o(t) \int_{\mathcal X}\!\Big[(\partial_t\lambda_i) f
+ \lambda_i(\partial_t f)\Big]\,dx  } 
\approx \frac{\partial_t\lambda_i}
{\displaystyle  \int_{\mathcal X}\! \Big[(\partial_t\lambda_i) f
+ \lambda_i(\partial_t f)\Big]\,dx} 
\approx \mathcal R_i(x,t) 
\end{equation}
where we in the last step used the the condition $\lambda_o \approx \bar{\lambda}_o(t) \Rightarrow \lambda_o /\bar{\lambda}_o(t) \approx 1$.

\clearpage 

\end{document}